\documentclass[10pt]{article}

\usepackage{amsmath}
\usepackage{amsthm}
\usepackage{amsfonts}
\usepackage{graphicx,psfrag,epsf}
\usepackage{enumerate}
\usepackage{natbib}
\usepackage{setspace}

\usepackage{amssymb}
\usepackage{url}

\let\emptyset\varnothing

\newcommand{\R}{\mathbb{R}}

\newcommand{\argmin}{\mathop{\mathrm{argmin}}}

\newcommand{\bbeta}{\mbox{\boldmath$\beta$}}
\newcommand{\boldbeta}{\mbox{\boldmath$\beta$}}
\newcommand{\Var}{\textrm{Var}}

\newcommand{\boldeta}{\mbox{\boldmath$\eta$}}
\newcommand{\boldpi}{\mbox{\boldmath$\pi$}}
\newcommand{\boldx}{\textbf{x}}
\newcommand{\boldA}{\textbf{A}}

\newcommand{\boldy}{\textbf{y}}
\newcommand{\boldd}{\textbf{d}}

\newcommand{\boldf}{\textbf{f}}
\newcommand{\boldg}{\textbf{g}}
\newcommand{\boldh}{\textbf{h}}
\newcommand{\boldv}{\textbf{v}}

\newcommand{\boldZero}{\textbf{0}}
\newcommand{\bolda}{\textbf{a}}
\newcommand{\boldD}{\textbf{D}}

\newcommand{\boldF}{\textbf{F}}
\newcommand{\boldb}{\textbf{b}}
\newcommand{\boldB}{\textbf{B}}
\newcommand{\boldM}{\textbf{M}}

\newcommand{\boldV}{\textbf{V}}
\newcommand{\boldW}{\textbf{W}}
\newcommand{\boldw}{\textbf{w}}
\newcommand{\boldY}{\textbf{Y}}
\newcommand{\boldC}{\textbf{C}}

\newcommand{\boldH}{\textbf{H}}

\newcommand{\bcompact}{ \def \baselinestretch{1.5} \large \normalsize }

\newcommand{\mc}{\multicolumn}

\newtheorem{theorem}{Theorem}
\newtheorem{corollary}{Corollary}

\begin{document}
\bcompact
\begin{small}
\begin{center}

\noindent {\LARGE \bf Multivariate Design of Experiments for Engineering Dimensional Analysis}

\vspace{.1in}

Daniel J. Eck

\vspace{.01in}

\textit{Department of Biostatistics, Yale University, New Haven, CT, United States}

\vspace{.1in}

R. Dennis Cook

\vspace{.01in}

\textit{School of Statistics, University of Minnesota, Minneapolis, MN, United States}

\vspace{.1in}

Christopher J. Nachtsheim

\vspace{.01in}

\textit{Carlson School of Management, University of Minnesota, Minneapolis, MN, United States}

\vspace{.1in}

Thomas A. Albrecht

\vspace{.01in}

\textit{Boston Scientific, Minnetonka, MN United States}
\end{center}
\end{small}

\singlespacing
\begin{abstract}

\noindent We consider the design of dimensional analysis experiments when there is more than a single response.  We first give a brief overview of dimensional analysis experiments and the dimensional analysis (DA) procedure.  The validity of the DA method for univariate responses was established by the Buckingham $\Pi$-Theorem in the early 20th century.  We extend the theorem to the multivariate case, develop basic criteria for multivariate design of DA and give guidelines for design construction. Finally, we illustrate the construction of designs for DA experiments for an example involving the design of a heat exchanger.\end {abstract}


\doublespacing


\doublespacing
\section{Introduction}

Dimensional analysis (DA) is a methodology developed in physics for reducing the number and complexity of experimental factors so that the relationship between the factors and the response can be determined efficiently and effectively.  DA is a powerful empirical technique for two reasons:

\begin{enumerate}
\item
Dimension reduction.  If a response appears to depend on $p$ physical predictors or factors, dimensional analysis can reduce the number of factors required to $p-k$ \textit{dimensionless} factors, where the reduction $k$ is often between one and four, and is usually given by the number of \textit{fundamental dimensions} in the problem.  With fewer factors, fewer runs are needed, and simpler designs can be employed for modeling the response.
\item Generalization of results.  Because the DA process converts all factors and responses to dimensionless quantities, the model can be used to predict results accurately, for values of the factors that are far outside the design region.  Thus, for example, an engineer designing propellers for a wind turbine can do the experimental work in a laboratory using prototype propellers whose lengths are only a few centimeters, and then generalize the results to the actual propeller sizes that currently can be as large as 82 meters.
\end{enumerate}

A compelling MIT instructional video\footnote{https://ocw.mit.edu/resources/res-tll-004-stem-concept-videos-fall-2013/videos/problem-solving/dimensional-analysis/} demonstrates how NASA used DA to compute the diameter of the parachute that would be required to slow the landing of the Mars rover to the desired velocity of 90m/s.   The response variable, velocity, can be modeled as a function of four independent variables: the diameter of the parachute, the mass of the rover, gravitational acceleration, and the density of the atmosphere.  There are three \textit{base quantities} involved in this formulation: length, $L$, mass, $M$, and time $T$.  A base quantity such as mass can be measured in different units, such as pounds or kilograms, but in either case the base quantity is mass.  A \textit{derived quantity of the first kind} is a quantity that is constructed from products of powers of base quantities.  Velocity is a derived quantity of the first kind because its dimension is $LT^{-1}$.  Following standard practice, we write $[v] = LT^{-1}$ to denote that the fundamental dimension of velocity is $LT^{-1}$.
\vspace{.1in}

\singlespacing
To summarize, for the rover example, we have:

\begin{equation}
v = f(d,m,g,\rho)
\label{MarsFunction}
\end{equation}
\noindent where:
\begin{align*}
v &= \text{velocity   (dependent variable)} &[v] &= L/T\\
d &= \text{diameter of parachute} &[d] &= L \\
m  &= \text{mass of the rover} &[m] &= M\\
g &= \text{gravitational acceleration} &[g] &= L/T^2\\
\rho &= \text{density of the atmosphere} &[\rho] &= M/L^3
\end{align*}
\doublespacing
\noindent  Since there are three fundamental dimensions involved---L (length), M (mass), and T (time)---the Buckingham $\Pi$ theorem (Buckingham, 1914), tells us that we can reduce the set of variables from five to $5-3=2$ dimensionless variables.  The DA model is $\pi_0 = \phi(\pi_1)$, or in terms of the variables:
\begin{equation}
\frac{v}{\sqrt{dg}} = \phi\left(\frac{\rho d^3}{m}\right)
\label{DAModel}
\end{equation}
\noindent The dimensionless variables $\pi_0$ and $\pi_1$ are sometimes referred to as the  ``$\pi$ groups.''  The terminology arose because the dimensionless variables are always products ($\Pi$) of powers of the original variables.  We describe a simple, step-by-step methodology for deriving the $\pi$ groups in Section 2.  At this point we know the dimensionless quantities, but we do not know the form of $\phi$.  Generally, a numerical or physical experiment is needed to estimate $\phi$.

\begin{figure}
\begin{center}
\includegraphics[width = 6in]{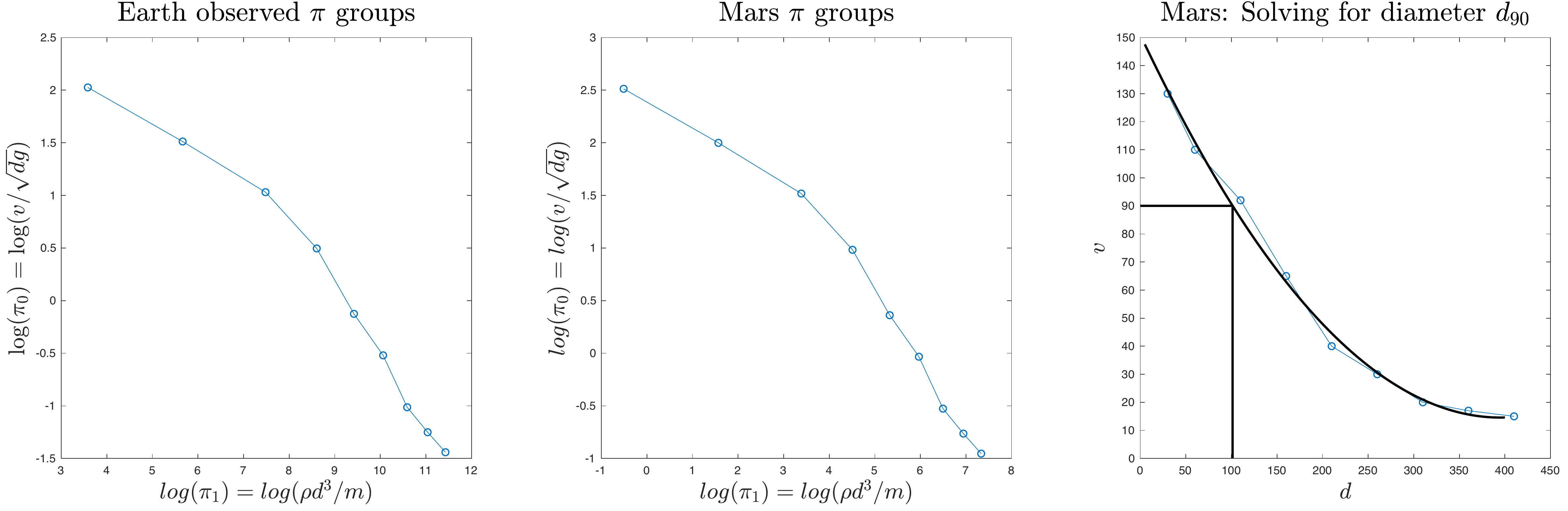}
\caption{Solving for the solution diameter, $d^*$.}
\label{MarsPlots}
\end{center}
\end{figure}

In order to estimate $\phi$, we will vary $\pi_1$ in a one-factor experimental design within a specified range, and record $\pi_0$.  To vary $\pi_1$, we need to vary $\frac{\rho d^3}{m}$.  This can be accomplished by varying only $d$, so that $\rho$, the density of the atmosphere, and $m$, the mass of the rover, need not be varied.  We then record values of the dependent variable $\pi_0 = \frac{v}{\sqrt{dg}}$ for each setting of $\pi_1$.  $\rho$ and $g$ will be set to the Earth's atmospheric density and gravitational acceleration, respectively, ($\rho = 1.225kg/m^3$ and $g= 9.8m/s^2$), and $d$ will be as specified in $\pi_1$, so that we really only need to record $v$.   A plot of simulated data (after taking logarithms) is shown in Figure \ref{MarsPlots}a. How can the model be used to determine the required parachute diameter $d^*$ that will slow the rover to 90 meters per second on Mars? We simply change the gravitational acceleration to the Mars version ($g = 3.7m/s^2$), change the atmospheric density to the Mars version ($\rho = 0.02kg/m^3$) and we now have a relation that works on Mars without having to perform a single experimental run external to the Earth's surface. Figure \ref{MarsPlots}b shows the values of $\pi_0$ and $\pi_1$ on Mars.  Finally, the value of the diameter $d^*$ (about $101m$) required to slow the rover to $v = 90 m/s$ on Mars is obtained from a quadratic smooth of $v = \pi_0\sqrt{dg}$ versus $d$ as shown in Figure \ref{MarsPlots}c.

The example clearly demonstrates the power of dimensional analysis experiments, both in terms of the kinds of dimension reduction that are possible, and in terms of the utility that results from the ability to extrapolate results beyond the experimental region. We note that physics is not the only area in which dimensional analysis has been successfully applied.  \citet[p. 296]{white} writes: ``Specialized books have been published on the application of dimensional analysis to metrology, astrophysics, economics, chemistry, hydrology, medications, clinical medicine, chemical processing pilot plants, social sciences, biomedical sciences, pharmacy, fractal geometry, and even the growth of plants.''

The benefits of the DA process do not come without some attendant complications. First, the DA model can be highly nonlinear. For this reason, an experimental design must be capable of estimating models  of higher order than those typically assumed in screening or response surface studies. Second, omission of a key explanatory variable can be fatal to the DA process. In an effort to alleviate this concern, \citet{albrecht}, proposed ``robust-DA'' designs that permit simultaneous estimation of the DA model and a standard empirical model in the original factors. The robust-DA approach maximizes the efficiency of the DA design in the dimensionless factors, subject to lower-bound constraint on the efficiency of the design for the original factors.  We emphasize that close collaboration between the statistician and the domain expert is critical to avoid the pitfall of excluding key explanatory variables from the analysis.

Although multiple responses are frequently present in DA experiments, design for multivariate responses in DA experiments has not been considered. In this paper, we extend the Buckingham $\Pi$-Theorem to the multivariate response case, give strategies for design of DA experiments for multiple responses, and illustrate results through a standard example.  A brief outline of the paper is as follows.  In Section 2, we provide a brief overview of the DA process.  The extension of the Buckingham $\Pi$-Theorem to multivariate responses is given in Section 3, and the design of experiments for multivariate DA problems is considered in Section 4.  Methods for constructing parametric and non-parametric designs are described in Section 5.  Finally, a real illustration involving the design of heat exchanger is provided in Section 6, and we conclude with a discussion in Section 7.

\section{Overview of DA \label{OverviewOfDA}}
In this section, we provide a brief overview of the DA process and we describe a step-by-step method due to \citet{ipsen} for deriving the DA model. For more detail, see, for example, \citet{sonin} and/or \citet{albrecht}.

When implementing DA, physical quantities are classified as either base quantities or derived quantities. Base quantities are composed of a single fundamental dimension. In physics, the System International (SI) states that length, mass, time, electrical current, temperature, amount, and luminous intensity are all base quantities. In economics or operations research the base quantity of cost is also of importance. A base quantity can be measured using different measurement systems. For example, one can use meters, feet, or inches to measure length. As noted in the introduction, a derived quantity of the first kind is a physical quantity that is comprised of a power-law combination of base quantities.

It has been shown that not all formulas can be used to represent physical quantities. Because base quantities all have a physical origin, the ratio of the measurements of any two base quantities does not change if the base unit changes. This is known as the \textit{principle of absolute significance} \citep*{brid}. The principle of absolute significance will hold for a physical quantity $x$ having a monomial formula only if it assumes the power-law form:
$$
  x = \gamma \prod_{i=1}^k Z_i^{b_i},
$$
where $Z_i$ is the numerical value of the $i$th base quantity and the coefficients $\gamma, b_1,..., b_k$ are real numbers. Thus, all physical quantities have power-law form and no other form represents a physical quantity. If the units chosen for the $i$th dimension are changed by a factor $c_i$, for $i = 1,..., k,$ it follows that the value of $x$ becomes $x^{\prime} = c^{-1}x$, where $c = \prod_{i=1}^k c_i^{b_{i.}}$. Finally, we say that a derived quantity is dimensionless if its value does not change when the units of the base quantities change. As an example, if we wish to calculate the velocity in meters per second of an object that has traveled 60 meters in 10 seconds, we would set $Z_1 = 60 m$, $b_1 = 1$, $Z_2 = 10 s$, $b_2 = -1$, and $\gamma = 1$, yielding $(1)(60 m)^1(10 s)^{-1} = 6 m/s$. We can then convert the same measurement to the unit system of kilometers ($c_1 = 1000 m/km$) per hour ($c_2 = 3600 s/h$) by calculating $c = (1000 m/km)^1(3600 s/h)^{-1} = 5 \textit{m-h}/18\textit{km-s}$ then applying the formula $x^{\prime} = c^{-1}x$ to yield: $x^{\prime} = (5\textit{m-h}/18\textit{km-s})^{-1}(6 m/s) = 21.6 km/h$.


There are three steps in a typical DA process:

\begin{enumerate}
\item Identify the dependent and $p$ independent variables.  In the Mars rover example of the Introduction, (\ref{MarsFunction}) stipulates that the dependent variable is $v$, and that the four independent variables are $d$, $m$, $g$ and $\rho$.  In what follows, we refer to $i$th independent variable as $x_i$, $i = 1,\ldots,p$, and the design space as $\chi$.
\item Identify the dimensionless dependent and independent $\pi$-groups.  These are denoted $\pi_0, \ldots, \pi_{p-k}$, and we use $\chi_{\pi}$ to denote the design space in the dimensionless independent variables.
\item Carry out a numerical or physical experiment to estimate the form of $\phi$ in the DA model: $\pi_0 = \phi(\pi_1, \ldots, \pi_{p-k})$.
\end{enumerate}

A simple, stepwise process due to \citet{ipsen} can be used to derive the dimensionless quantities.   In essence, at each step, an independent variable is chosen to eliminate one of the dimensions from the model.  The process terminates when all remaining variables are dimensionless.  We illustrate the \citet{ipsen} process now in the context of the Mars rover example (\ref{MarsFunction}), although we note that other, equivalent methods can be used.  See, e.g., \citet{white}, or \citet{sonin}.

The step-by-step method begins, in Step 0 with a listing the variables and their dimensions in adjacent columns, as shown in the left panel of Table \ref{moonDAModel}.  In Step 1, we have arbitrarily chosen to eliminate the length dimension $L$ using the diameter variable $d$.  This means that we multiply each of the variables whose dimension includes length ($L$) by an appropriate power of $d$ to eliminate $L$.  There are three variables that are a function of $L$ (besides $d$), namely $v$, $g$ and $\rho$. We first multiply $v$ by $d^1$ creating a new variable, $vd$ whose dimension is $T^{-1}$. Similarly we multiply $g$ by $d^{-1}$ and $\rho$ by $d^3$ yielding new variables $g/d$ and $\rho d^3$ whose dimensions are $T^{-2}$ and $M$, respectively. At this point we create two new columns (labeled ``Step 1 Result'') containing the new variables and their dimensions.  We draw a horizontal bar in the ``$d$'' row, because this variable has been eliminated (e.g., incorporated into the other variables), and we are ready for Step 2.

In Step 2, we have (arbitrarily) chosen to eliminate the mass dimension ($M$) from the Step 1 columns using the mass of the rover, $m$. Aside from $m$, there is only one variable, $\rho d^3$ whose dimension is a function of $M$.  Since that dimension of $\rho d^3$ is exactly $M$ we multiply $\rho d^3$ by $m^{-1}$ giving $\rho d^3/m$, which is dimensionless, and so signified using a ``1.''  We can now move the new (and existing) variables and their dimensions into a new column labeled ``Step 2 Result.''  Since $m$ has been eliminated as a separate variable, a horizontal bar is listed in that row.

In Step 3, there is only one dimension left ($T$) and only one independent variable ($gd^{-1}$) that can be used to eliminate $T$ in the column labeled ```Step 2 Result.''  We multiply  the response variable $vd^{-1}$ by $1/\sqrt{gd^{-1}}$ giving $v/\sqrt{dg}$, which is dimensionless.  Moving the two dimensionless variables to the final two columns labeled ``Step 3'' and adding a horizontal bar to the $g$ row completes the process.  The process terminates at this point because all remaining variables are dimensionless.  The DA model has $\pi_0 = v/\sqrt{dg}$ and $\pi_1=\rho d^3/m$ as in (\ref{DAModel}).

\begin{table}
\begin{center}
\begin{tabular}{|ll|cc|cc|cc|}
\hline
&&\mc{2}{|c|}{Step 1 Result:} & \mc{2}{|c|}{Step 2 Result:} & \mc{2}{|c|}{Step 3 Result:} \\
&& \mc{2}{|c|}{Remove $L$ from} & \mc{2}{|c|}{Remove $M$ from} & \mc{2}{|c|}{Remove $T$ from} \\
\mc{2}{|c|}{Step 0: Initialize}& \mc{2}{|c|}{Step 0 using $d$} & \mc{2}{|c|}{Step 1 using $m$} & \mc{2}{|c|}{Step 2 using $gd^{-1}$} \\
\hline
Variable & Dimension & Var. & Dim. & Var. & Dim. & Var. & Dim. \\
\hline
$v$	& $LT^{-1}$ & $vd^{-1}$ & $T^{-1}$ & $vd^{-1}$ & $T^{-1}$ & $v/\sqrt{dg}$ & $1$ \\
$d$	& $L$ & \mc{2}{|c|}{\text{--------------}}  & \mc{2}{|c|}{\text{--------------}}  & \mc{2}{|c|}{\text{--------------}}\\
$m$ & $M$ & $m$ & $M$ & \mc{2}{|c|}{\text{--------------}} & \mc{2}{|c|}{\text{--------------}} \\
$g$	& $LT^{-2}$ & $gd^{-1}$ & $T^{-2}$ & $gd^{-1}$ & $T^{-2}$ & \mc{2}{|c|}{\text{--------------}} \\
$\rho$ & $ML^{-3}$ & $\rho d^3$ & $M$ & $\rho d^3m^{-1}$ & $1$  & $\rho d^3m^{-1}$ & $1$  \\
\hline
\end{tabular}
\caption{Deriving the DA model for the Mars rover example}
\label{moonDAModel}
\end{center}
\end{table}

Two comments are in order:
\begin{enumerate}
\item As noted, the order in which the fundamental dimensions are eliminated is arbitrary; different orders can lead to different $\pi$ groups and therefore different DA models.  Engineers and scientists generally prefer $\pi$ groups that correspond to well-known dimensionless quantities (see, e.g., White, 1999, for a list of such quantities in fluid mechanics).  In the Mars rover example, $\pi_0$ is is the Froude number, which is the ratio of a body's energy to inertia due to gravitational forces.  Also, $\pi_1$ is the square root of the inverse of the Richardson number, which is the ratio of potential to kinetic energy.
\item The size of the dimension reduction $k$ can be less than the number fundamental dimensions in some situations.  This occurs, for example, whenever two or more dimensions are eliminated simultaneously in any step of the the Ipsen (1960)  procedure.
\end{enumerate}

We now take up the the design of DA experiments in the presence of multiple responses, starting with a generalization of the Buckingham $\Pi$-Theorem.

\section{Buckingham $\Pi$-Theorem for multivariate responses \label{MVBuckingham}}
The examples in Albrecht, et al. (2013) show that DA is a valuable tool that
provides dimension reduction of the predictors when the response is a
scalar. The same ideas apply to any regression or design of experiments
problem with a vector-valued response. In this setting, the Buckingham
$\Pi$-Theorem has a multivariate analog where $\boldY \in \R^r$ is the
vector of responses and $\boldx \in \R^p$ is the vector of predictors.
Both types of variables can be expressed as power-law combinations of
$k$ fundamental dimensions that are measured with respect to a
particular system of units. Let
$\boldb_i = (b_{1i},..., b_{ki})^{\prime}$,$i = 1,...,p$ be the
dimension vector of $x_i$ where $b_{ji} \in \R$ is the power to which
the $j$th fundamental dimension is raised in $[x_i]$. Define
$$
  \boldB = \left( \begin{array}{cccc}
    b_{11} & b_{12} & \cdots & b_{1p} \\
    b_{21} & b_{22} & \cdots & b_{2p} \\
     \vdots  & \vdots  &  & \vdots \\
    b_{k1} & b_{k2} & \cdots & b_{kp}
  \end{array} \right)
$$
to be the $k \times p$ dimension matrix for the predictors in a given problem.
Let $\bolda_i = (a_{1i},..., a_{ki})^{\prime}$, $i = 1,...,r$, be the
dimension vector of $Y_i$ where $a_{ji} \in \R$ is the power to which the
$j$th fundamental dimension is raised in $[Y_i]$. Define
$$
  \boldA = \left( \begin{array}{cccc}
    a_{11} & a_{12} & \cdots & a_{1r} \\
    a_{21} & a_{22} & \cdots & a_{2r} \\
    \vdots  & \vdots  &  & \vdots \\
    a_{k1} & a_{k2} & \cdots & a_{kr}
  \end{array} \right)
$$
to be the $k \times r$ dimension matrix for the response variables in a given
problem.
The Multivariate Buckingham $\Pi$-Theorem assumes the following where the
assumptions and some of the theoretical details follow from
\citet*[p. 5-9]{bluman}.

\begin{theorem} \label{MVtheorem}
Assume the following:
\begin{itemize}
\item[(i)] A vector $\boldY \in \mathbb{R}^r$ has a functional relationship
  with $p$ predictors $(x_1,...,x_p)$:
  \begin{equation}\label{functional-form1}
    \boldY = \boldf(x_1,...,x_p),
  \end{equation}
  where $\boldf$ is an unknown function of the predictors.
\item[(ii)] The quantities $(Y_1,...,Y_r,x_1,...,x_p)$ involve $k$ fundamental
  dimensions labeled by $L_1,..., L_k$. Then it is assumed that
  $\boldA \subseteq$ span$(\boldB)$. 
\item[(iii)] Let $Z$ represent any of $(Y_1,...,Y_r,x_1,...,x_p)$. Then,
  $$
    \left[Z\right] = \prod\limits_{i=1}^k L_i^{\alpha_i}
  $$
  for some $\alpha_i \in \mathbb{R}, i = 1,..,k$ which are the dimension
  exponents of $Z$.
\item[(iv)] For any set of fundamental dimensions one can choose a system of
  units for measuring the value of any quantity $Z$. A change from one system
  of units to another involves a positive scaling of each fundamental
  dimension which in turn induces a scaling of each quantity $Z$. Under a
  change of system of units the value of a dimensionless quantity is
  unchanged, i.e. its value is invariant under an arbitrary scaling of
  fundamental dimension.
\end{itemize}


Assumptions (i)-(iv) give:


\begin{enumerate}
\item[(i)] Formula (\ref{functional-form1}) can be written in terms of
  dimensionless quantities.
\item[(ii)] The number of dimensionless predictors is
  $p - \text{rank}(\boldB)$ where $\text{rank}(\boldB)$ is the rank of the
  matrix $\boldB$.
\item[(iii)] Let $x_i = (\pi_{1i},..., \pi_{pi})^{\prime}$,
  $i = 1,...,p - \text{rank}(\boldB)$ represent the linearly independent
  solutions of the system $\boldB x_i = 0$. Let
  $\bolda_i = (a_{1i},..., a_{ki})^{\prime}$ be the dimension vector for
  response $Y_i, i =\, 1,...,r$ and let
  $\boldy_i = (\rho_{1i},...,\rho_{pi})$ represent a solution to the
  system $\boldB\boldy_i = -\bolda_i$. Then formula (\ref{functional-form1})
  simplifies to $\tilde{\boldpi} = \boldh(\pi_1,...,\pi_{p - \text{rank}(\boldB)})$
  where $\tilde{\boldpi} \in \R^r$. All elements of $\tilde{\boldpi}$ and
  $\pi_i$ are dimensionless quantities for all
  $i = 1,...,p - \text{rank}(\boldB)$.
\end{enumerate}
\end{theorem}

The proof of the Multivariate Buckingham $\Pi$-Theorem is included in
Appendix 1. This Theorem follows from the univariate Buckingham $\Pi$-Theorem
applied to each response variable provided that
$\boldA \subseteq$ span$(\boldB)$ holds.
To see why $\boldA \subseteq$ span$(\boldB)$ is needed,
consider a design problem with two responses and three predictors where each
variable has fundamental dimensions given by
$\left[Y_1\right] = ML$, $\left[Y_2\right] = MT$, $\left[x_1\right] = ML$,
$\left[x_2\right] = MT^2$, $\left[x_3\right] = MT^2$.
In this setup
$$
  \boldA = \left(\begin{array}{cc}
  1 & 1 \\
  1 & 0 \\
  0 & 1
  \end{array}\right), \qquad \boldB = \left(\begin{array}{ccc}
  1 & 1 & 1 \\
  1 & 0 & 0 \\
  0 & 2 & 2
  \end{array}\right).
$$
We use DA to create a single dimensionless predictor $x_2/x_3$ and a
dimensionless response $Y_1/x_1$. However, no combination of predictors can
combine with $Y_2$ to yield a second dimensionless response. This is a result
of violating $\boldA \subseteq$ span$(\boldB)$.

To see why the reduction of predictors is by $\text{rank}(\boldB)$ and not
$k$, consider the following example. Suppose there are two responses and
three predictors where each has fundamental dimensions given by
$\left[Y_1\right] = M$, $\left[Y_2\right] = LT^{-1}$, $\left[x_1\right] = M$,
$\left[x_2\right] = LT^{-1}$, $\left[x_3\right] = LT^{-1}$.
In this setup
$$
  \boldA = \left(\begin{array}{cc}
  1 &  1 \\
  0 &  1 \\
  0 & -1
  \end{array}\right), \qquad \boldB = \left(\begin{array}{ccc}
  1 &  0 &  0 \\
  0 &  1 &  1 \\
  0 & -1 & -1
  \end{array}\right).
$$
We see that $\boldA \subseteq$ span$(\boldB)$ holds. Using DA, we
create a single dimensionless predictor $x_2/x_3$ and two dimensionless
responses $Y_1/x_1$ and $Y_2/(x_1x_3)$. In this example there are $k = 3$
fundamental dimensions and $\text{rank}(B) = 2$. Our DA model consists of
$p - \text{rank}(B) = 1$ dimensionless predictor.

When $\boldA \subset \text{span}(\boldB)$ the Multivariate
Buckingham $\Pi$-Theorem holds and Dimensional Analysis is applicable for
multivariate models. However, all is not lost when
$\boldA \setminus \text{span}(\boldB) \neq \emptyset$.
When $\boldA \setminus \text{span}(\boldB) \neq \emptyset$,
it may be the case that certain responses need to be excluded from the
DA model. For $j = 1,...,r$ let $\boldA_j$ denote the $j^{th}$ column of
$\boldA$ and let $\boldA_{-j}$ be the matrix $\boldA$ with $\boldA_j$ removed.
Suppose that
$
  \boldA_j \notin \text{span}(\boldA_{-j}, \boldB)
$
then the response $\boldY_j$ cannot be made to be dimensionless and cannot
be used to make other responses dimensionless. Therefore $\boldY_j$ must be
excluded from consideration in the DA model.
Thus we require that $\boldA_j \in$ span$(\boldA_{-j},\boldB)$. This means
that maybe some responses cannot be represented by either other
responses alone or the predictors alone, but may be represented by the
combination of some responses and predictors.
With such cases in mind we proceed with the a corollary to the Multivariate
Buckingham $\Pi$-Theorem that accounts for when
$
  \boldA \setminus \text{span}(\boldB) \neq \emptyset.
$

\begin{corollary}
Multivariate Buckingham $\Pi$-Theorem II. Suppose that $\boldA \setminus \text{span}(\boldB) \neq \emptyset$ and exclude responses such that $\boldA_j \notin \text{span}(\boldA_{-j}, \boldB)$ from consideration. Suppose that $0 < r_1 \leq r$ responses remain. Let $r_2$ be the number of responses not belonging to $\text{span}(\boldB)$, let $\boldA^{\textstyle{*}}$ be the dimension matrix corresponding to these responses and put $\boldC = \left[\boldA^{\textstyle{*}} \; \boldB\right]$. Assume the following:
\begin{enumerate}
\item[(i)] A vector $\boldY \in \mathbb{R}^{r_1}$ has a functional relationship with $p$ predictors $(x_1,...,x_p)$ given by $\boldY = \boldf(x_1,...,x_p)$
where $\boldf$ is an unknown function of the predictors.
\item[(ii)] The quantities $(Y_1,...,Y_r,x_1,...,x_p)$ involve $k$ fundamental dimensions labeled by $L_1,..., L_k$ where $k < p$ is assumed to ensure a meaningful problem.
\item[(iii)] Let conditions (iii)-(iv) be as in the Multivariate Buckingham $\Pi$-Theorem.
\end{enumerate}
These assumptions give:
\begin{itemize}
\item[(i)] The number of dimensionless predictors is $p - \text{rank}(\boldB)$.
\item[(ii)] The number of dimensionless response variables is $r_3 = r_1 - \text{rank}(\boldC) + \text{rank}(\boldB)$.
\item[(iii)] There exists a function $\boldg:\R^{r_1} \to \R^{r_3}$ such that
\begin{equation} \label{functional-form2}
  \boldY^{\prime} = \boldg(\boldY) = \boldg \circ \boldf(x_1,...,x_p)
\end{equation}
can be written in dimensionless quantities.
\end{itemize}
\end{corollary}

This Corollary corresponds to a two-stage framework to do DA.
Model~\eqref{functional-form2} indicates that you can reduce the responses not to
dimensionless, but to those that can be represented by the predictors.
The proof of this Corollary and an example of this setting are both included
in Appendix 1.

The steps required to implement a DA model when the response is multivariate
are similar to those outlined in Section 2.  The only
difference is that the condition $\boldA \subseteq$ span$(\boldB)$ is required
for the DA procedure to yield a set of dimensionless variables.

We consult \citet*[p. 722]{white} for an example of dimensional analysis in
the presence of multiple responses. For a given pump design, the output
variables $gH$ and brake horsepower $(bhp)$ should be dependent upon
discharge $Q$, impeller diameter $D$, and shaft speed $s$, at least. Other
possible parameters include fluid density $\rho$, viscosity $\mu$, and
surface roughness $\epsilon$. Thus, we have a functional relation where
$\boldf:\R^6 \to \R^2$ given by

\begin{equation} \label{eqn}
  \left(\begin{array}{c}
      gH \\
      bhp
    \end{array}\right) = \boldf(Q,D,s,\rho,\mu,\epsilon)
\end{equation}
where the variables are comprised of dimensions as seen in the table below:

\begin{center}
\begin{tabular}{|l||c|}
\hline
variable & dimensions \\
\hline
$gH$   & $\left[L^2T^{-2}\right]$ \\
$bhp$ & $\left[ML^2T^{-3}\right]$ \\
\hline
$Q$        & $\left[L^3T^{-1}\right]$ \\
$D$        & $\left[L\right]$ \\
$s$        & $\left[T^{-1}\right]$ \\
$\rho$     & $\left[ML^{-3}\right]$ \\
$\mu$      & $\left[ML^{-1}T^{-1}\right]$ \\
$\epsilon$ & $\left[L\right]$ \\
\hline
\end{tabular}
\end{center}

There are eight variables in this model and a total of three fundamental
dimensions, length ($L$), mass ($M$) and time ($T$). In this example we see
that $\boldA \subseteq \text{span}(\boldB)$. Therefore we can
express the functional relationship (\ref{eqn}) in terms of three
dimensionless quantities as a result of the Multivariate Buckingham
$\Pi$-Theorem. Implementation with respect to this example is continued in
Section 5.

\section{Design for DA with multiple responses}
In this section, we consider the design of DA experiments for multiple responses.  We assume that the DA model has been formulated, that there are $r$ responses, $\boldY = (Y_1,...,Y_r)^{\prime}$, $p$ dimensionless factors, $\boldx = (x_1, \ldots, x_p)^{\prime}$, so that our DA model can be written:
$$
  \textrm{E}(\boldy|\boldx) = \boldh(\boldx) = \left(
  \begin{array}{c}
    h_1(\boldx)\\
    \vdots \\
    h_r(\boldx)
  \end{array} \right)
$$
In the univariate setting, when the form of the DA model $h$ is unknown and potentially complex, Albrecht et al. (2013) identified the use of a nonparametric \textit{uniform} design as one alternative.  In a uniform design, the design points are distributed in such a way that the empirical cumulative distribution is as close as possible to the cumulative distribution of a uniform probability measure on the design space.  We note that for nonparametric designs, the multivariate design will be the same as the univariate design for any one of the responses provided that predictors are defined on a product space.  Thus, given the multivariate DA model, there are no new design issues.

The alternative approach suggested by \citet{albrecht} is to design for estimation of third- or higher-order polynomials in the dimensionless factors, and they advocated the use of D-optimal designs in that context. They also suggested that the integrated variance might be more appropriate for design of dimensional analysis experiments, since the objective is to predict the expected response over the design space.  In this paper, I-optimality will be the primary design criterion of interest when polynomial models are to be estimated.

We assume that the design, denoted $\xi_n$, is exact and concentrated on the $n$ design points $\boldx_1, \ldots, \boldx_n  \in \R^p$. The value of the $j$th response variable for the $i$th run of the experiment can be modeled as:
\begin{equation} \label{model}
  y_j(\boldx_i) = \boldg_j^{\prime}(\boldx_i)\bbeta_j + \varepsilon_{ij}, \textrm{  for } i = 1,\ldots,n \textrm{  and } j = 1,\ldots,r
\end{equation}
where the model vectors $\boldg_j(\boldx)$, $j = 1,\ldots,r$, are known and the coefficient vectors $\bbeta$ are unknown. Let $\boldy_i$ denote the $r \times 1$ vector of responses for the $i$th run.  The multivariate formulation of model \eqref{model} is constructed with
$$
  \boldbeta = \left(
    \begin{array}{c}
    \boldbeta_1 \\
    \vdots \\
    \boldbeta_r \\
  \end{array}\right)
  \quad
  \boldf_j = \left(
    \begin{array}{c}
    \boldZero_{j,1} \\
    \boldg_j(\boldx) \\
    \boldZero_{j,2}
    \end{array}\right)
$$
for $j = 1,\ldots,r$ where $\boldZero_{j,1} \in \R^{m_1 + \cdots + m_{j-1}}$, $\boldZero_{j,2} \in \R^{m_{j+1} + \cdots + m_r}$, $\bbeta_j \in \R^{m_j}$, $\boldg_j(\boldx) \in \R^{m_j}$, $m_{\cdot} = m_1 + \cdots + m_r$, and $\boldbeta_i$ and $\boldbeta_j$, $i \neq j$, do not have terms in common. Here it is possible that $\boldf_i$ and $\boldf_j$, $i \neq j$, may have terms in common, but there is no reason to expect the regression coefficients of the common terms to be the same. The covariance matrix of the response vector is denoted
$$
  \Var(\boldy|\boldx) = \boldW^{-1}(\boldx)
$$
\noindent where $\boldW$(\boldx) is the weight matrix at $\boldx$. Let $\boldF(\boldx)$ denote the $m \times r$ matrix $[\boldf_1(\boldx),\boldf_2(\boldx),\ldots,\boldf_r(\boldx)]$.

Following, for example, Fedorov (1972), the best linear estimator for $\bbeta$ is
$$
  \hat{\bbeta} = \boldM^{-1}(\xi_n)\boldY
$$
\noindent where:
\begin{equation}
  \boldM(\xi_n) = \sum_{i=1}^n\boldF(\boldx_i)\boldW(\boldx_i)\boldF^{\prime}(\boldx_i) \qquad \textrm{and} \qquad \boldY = \sum_{i=1}^n\boldF(\boldx_i)\boldW(\boldx_i)\boldy_i
\label{InfoMatrix}
\end{equation}
\noindent The variance-covariance matrix of the estimator is
$$
  \boldD(\hat{\bbeta}) = \boldM^{-1}(\xi_n).
$$

\noindent Moreover, the best linear unbiased estimate of $\boldf_l^{\prime}(\boldx)\bbeta$, for $l = 1,\ldots,r$ is the function
\begin{equation}
  \hat{\eta}_l(\boldx) = \boldf_l^{\prime}(\boldx)\hat{\bbeta},
\label{BLUE}
\end{equation}

\noindent Let $\hat{\boldeta}(\boldx) = [\hat{\boldeta}_1(\boldx),\hat{\boldeta}_2(\boldx),\ldots,\hat{\boldeta}_r(\boldx)]^{\prime}$.  The dispersion matrix of $\hat{\boldeta}(\boldx)$ is
\begin{equation*}
  \boldd(\boldx,\xi_n) = \boldF^{\prime}(\boldx)\boldM^{-1}(\xi_n)\boldF(\boldx).
\end{equation*}

Clearly, from (\ref{InfoMatrix}), knowledge regarding the form of $\boldW(\boldx)$ is required for design construction.  Because all dimensionless factors are included in each model there should be no latent extrinsic factor that might induce a correlation between the responses.   In consequence, the errors can be reasonably taken to be uncorrelated.  We make the further simplifying assumption that the error variance matrix is constant over the design space.  That is $\Var(\boldy|\boldx) = \boldW^{-1}$. 
More specifically, we have:
$$
  \Var(\boldy|\boldx) =
  \begin{bmatrix}
      \sigma_1^2 & & \\
    & \ddots & \\
    & & \sigma_r^2
  \end{bmatrix} \textrm{ so that } \boldW =
  \begin{bmatrix}
    w_1 & & \\
    & \ddots & \\
    & & w_r
  \end{bmatrix}
$$
\noindent where $w_i = \sigma_i^{-2}$ for $i = 1,\ldots,r$.  We do not strictly assume that the $\{w_i\}$ are known, although this may at times be reasonable, since the variability of engineering measurement instruments is often known. Instead, by using relative efficiency, it is often possible to choose a design that performs well across various reasonable configurations of the component variances so that selecting a particular configuration becomes unnecessary.  This is indeed the case for the heat-exchanger example of Section 6.

Given this setup, from (\ref{InfoMatrix}) we have:
\begin{align}
\begin{split}
  \boldM(\xi_n) &= \sum\boldF(\boldx_i)\boldW\boldF^{\prime}(\boldx_i) \\
    &= \begin{bmatrix}
    w_1\boldM_1(\xi_n) & & \\
    & \ddots & \\
    & & w_r\boldM_r(\xi_n)
  \end{bmatrix}
  \label{M_i}
\end{split}
\end{align}
\noindent where $\boldM_i(\xi_n) = \sum_{j = 1}^n\boldf_i(\boldx_j)\boldf^{\prime}_i(\boldx_j)$.  One measure of the ``goodness'' of the design $\xi$ is given by the $D$ criterion:
$$
|\boldM(\xi_n)| = \prod_{i = 1}^r w_i|\boldM_i(\xi_n)|
$$
If the $r$ approximating models are identical such that $\boldf_1 = \cdots = \boldf_r$, the $D$ criterion simplifies to:
\[
|\boldM(\xi_n)| = (\prod_{i = 1}^r w_i)|\boldM_1(\xi_n)|^r
\]
\noindent Thus, the D-optimal design maximizes $|\boldM_1(\xi_n)|$.  As noted, our emphasis herein will be on the minimization of the integrated variance of prediction, that is, the $I$ criterion.  We have from (\ref{M_i})
\begin{equation}
\boldM^{-1}(\xi_n) = \boldD(\xi_n) =
\begin{bmatrix}
    w_1^{-1}\boldD_1(\xi_n) & & \\
    & \ddots & \\
    & & w_r^{-1}\boldD_r(\xi_n)
\end{bmatrix}
\label{Dmatrix}
\end{equation} \noindent where $\boldD_i(\xi_n) = \boldM_i^{-1}(\xi_n)$, for $i=1,\ldots,r$.
\noindent Let $v_{\chi} = \int_{\chi}d\boldx$ denote the volume of the design space $\chi$.  Then the average value of the dispersion matrix $\hat{\boldeta}(\boldx)$ over the design space is:
\begin{align*}
v_{\chi}^{-1}\int_{\chi}\boldd(\boldx,\xi_n)d\boldx &= v_{\chi}^{-1}\int_{\chi}\boldF^{\prime}(\boldx)\boldD(\xi_n)\boldF(\boldx)d\boldx \\
&= v_{\chi}^{-1}
\begin{bmatrix}
    w_1^{-1}\int_{\chi}\boldf_1(\boldx)\boldD_1(\xi_n)\boldf_1(\boldx)d\boldx & & \\
    & \ddots & \\
    & & w_r^{-1}\int_{\chi}\boldf_r(\boldx)\boldD_r(\xi_n)\boldf_r(\boldx)d\boldx
\end{bmatrix}
\end{align*}
\noindent Since $\int_{\chi}\boldf_i(\boldx)\boldD_i(\xi_n)\boldf_i(\boldx)d\boldx = \textrm{Trace}[\boldD_i(\xi_n)\int_{\chi}\boldf_i(\boldx)\boldf_i^{\prime}(\boldx)d\boldx] = \textrm{Trace}[\boldD_i\boldM_i]$, where $\boldM_i = \int_{\chi}\boldf_i(\boldx)\boldf_i^{\prime}(\boldx)d\boldx$, we have:
\[
v_{\chi}^{-1}\int_{\chi}\boldd(\boldx,\xi_n)d\boldx = v_{\chi}^{-1}
\begin{bmatrix}
    w_1^{-1}\textrm{Trace}[\boldD_1\boldM_1]& & \\
    & \ddots & \\
    & & w_r^{-1}\textrm{Trace}[\boldD_r\boldM_r]
\end{bmatrix}
\]
\noindent At this point the criterion is multivariate.  One natural way to reduce the criterion to a scalar is obtained by averaging.  Let
\begin{equation}
I_{MV}(\xi_n) = r^{-1}v_{\chi}^{-1}\sum_{i=1}^rw_i^{-1}\textrm{Trace}[\boldD_i\boldM_i]
\label{MVCriterion}
\end{equation}
If the forms of the $r$ approximating polynomials are identical, the criterion reduces to the minimization of the Trace$[\boldD_1\boldM_1]$.

\section{Guidelines for design construction \label{DesignIssues}}

In this section we use the pump example of Section \ref{MVBuckingham} to illustrate various approaches to design.  We consider both optimal design construction using parametric models and the construction of uniform nonparametric designs.


\subsection{Optimal design using parametric models \label{OptimalDesignGuidelines}}
As noted in the introduction, Albrecht et al., advocated the use of third- or higher-order models for estimating response surfaces in the dimensionless variables.  In their discussion, they also suggested the use of the model-robust ($\bar{L}$-optimal) designs of Cook and Nachtsheim (1983).  This approach can be easily combined with our multivariate criterion (\ref{MVCriterion}) leading to alternative design criteria.  Given a design criterion, there are alternative numerical approaches that can be employed for design construction.  In this section we explore these alternatives.  We do so in the context of the pump example of Section \ref{MVBuckingham}.  A more complicated example is explored in Section {\ref{HeatTransferExample}}.

We can rewrite (\ref{eqn}) as

\begin{equation} \label{eqn2}
  \left(\begin{array}{c}
      \frac{gH}{s^2D^2} \\
      \frac{bhp}{\rho s^3D^5}
    \end{array}\right) = \boldg\left(\frac{Q}{sD^3},\frac{\rho sD^2}{\mu},\frac{\epsilon}{D}\right)
\end{equation}

\noindent where $\frac{\rho sD^2}{\mu}$ and $\frac{\epsilon}{D}$ are recognized as the Reynolds number and roughness ratio respectively. Three new pump parameters have arisen: (1) the capacity coefficient,  $C_Q = \frac{Q}{sD^3}$, (2) the head coefficient $C_H = \frac{gH}{s^2D^2}$, and the power coefficient $C_P = \frac{bhp}{\rho s^3D^5}$.

For purposes of illustration, we make the simplifying assumption that the pump is being designed for use in one fluid only (e.g., water) and that roughness ratio is constant.  Thus $\epsilon$, $\mu$, and $\rho$ are constant.  The response models become:

\begin{equation} \label{eqn3}
  \left(\begin{array}{c}
      \frac{gH}{s^2D^2} \\
      \frac{bhp}{\rho s^3D^5}
    \end{array}\right) = \boldg\left(\frac{Q}{sD^3},\frac{\rho sD^2}{\mu}\right)
\end{equation}

\noindent where $\boldg:\R^2 \to \R^2$.  Expression (\ref{eqn3}) is a valid dimensionless functional form by the Multivariate Buckingham $\Pi$-Theorem since the set of fundamental dimensions present in the response variables is equal to the set of fundamental dimensions present in the predictors. The dimensionless variables are $\pi_1 = Q/(sD^2)$ and $\pi_2 = \rho s D^2/\mu$ . Since we are holding $\rho$ and $\mu$ constant, we can simplify the notation for the formula for $\pi_2$ to $\pi_2 = sD^2$ for this design region. The design region for the original variables $Q$, $s$, and $D$ is then:

\[  \chi = \{\, (Q,s,D): 4\leq Q\leq 30,\; 710\leq s\leq 1170,\; 28\leq D\leq 42 \,\}.  \]

\begin{figure}
\begin{center}
\includegraphics[width = 6in]{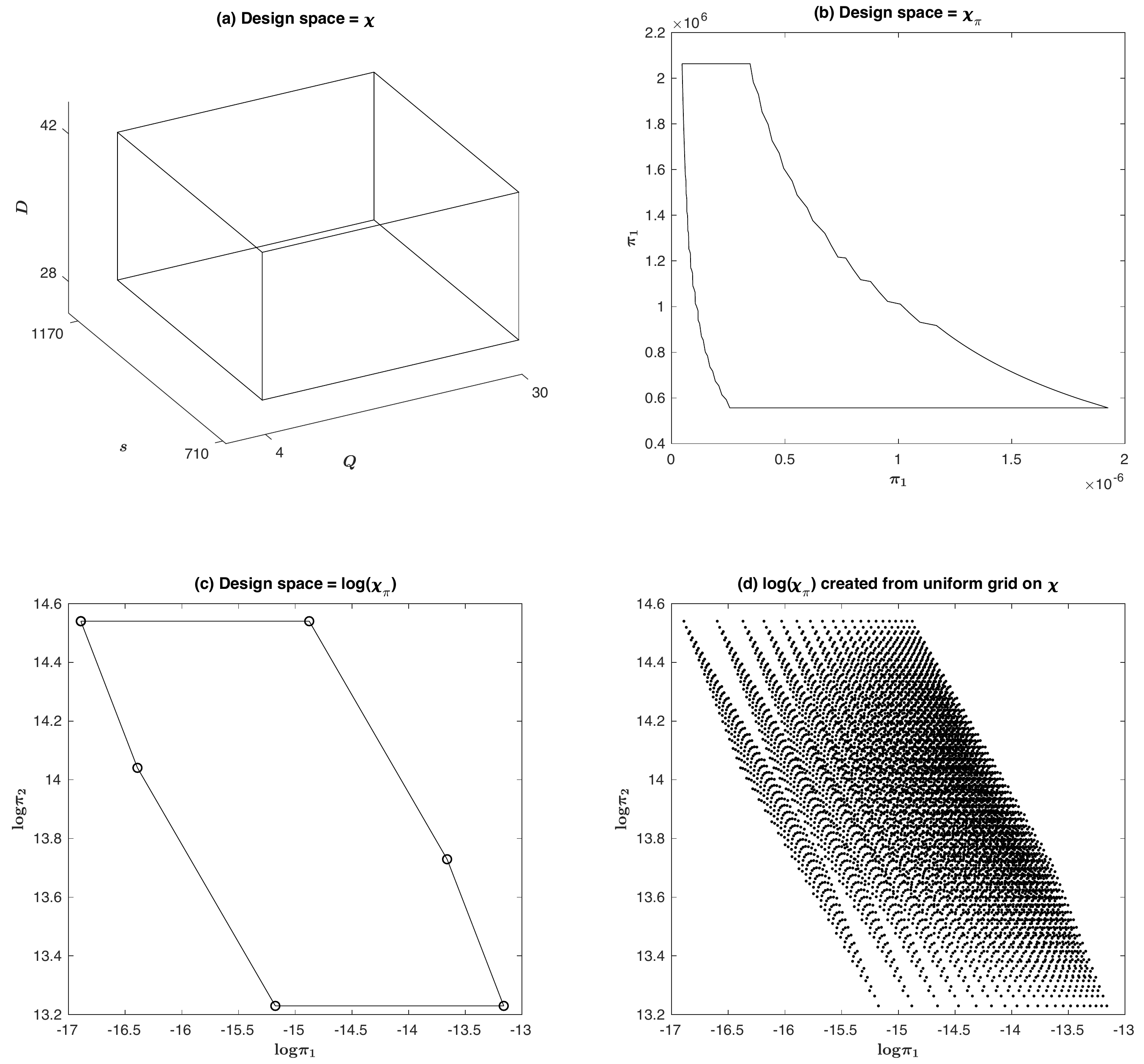}
\caption{Four design spaces for the pump design example.}
\label{2DAnd3DDesignSpaces}
\end{center}
\end{figure}

\noindent $\chi_{\pi}$ is shown in Figure \ref{2DAnd3DDesignSpaces}(a).  The dimensionless variables are $\pi_1 = \frac{Q}{sD^3} \quad \text{and} \quad \pi_2 = sD^2$. The design region corresponding to the dimensionless $\pi$-variables is given by
\[ \chi_{\pi} = \{\, (\pi_1, \pi_2): \pi_1 = Q/(sD^3), \pi_2 = sD^2 \; \text{where} \; (Q,s,D) \in \chi \,\}.  \]
\noindent $\chi_{\pi}$ is shown in Figure~\ref{2DAnd3DDesignSpaces}(b).  Albrecht et al., (2013) recommended working with the log of the $\chi_{\pi}$ design space, which, for this example is pictured in Figure~\ref{2DAnd3DDesignSpaces}(c).

Assume that we will employ a full fourth-order polynomial to approximate the response surface in the two $\pi$ groups.  We have three options, corresponding to the three design spaces, $\chi$,  $\chi_{\pi}$, and  log$\chi_{\pi}$, shown respectively in Figure \ref{2DAnd3DDesignSpaces}, panels (a), (b) and (c).  We examine each of these design approaches in turn.

\subsubsection{Search the $\chi$ space.}
In this case, we use an exchange algorithm to search for a ``best'' set of triples $\{(Q_i, n_i, D_i), i = 1,\ldots, n\}$ in $\chi$ that lead to a highly efficient design in $\chi_{\pi}$ or log$\chi_{\pi}$.  Within the algorithm, in order to evaluate the design criterion, points in $\chi$ are first mapped into points in $\chi_{\pi}$ or log$\chi_{\pi}$, and then the criterion (\ref{MVCriterion}) is evaluated for the resulting design.  An advantage to this approach is that the coordinate exchange algorithm can be easily implemented since $\chi$ is a hyper rectangle.  Generally, the coordinate-wise searches are performed on a uniform grid of (say) ten or more points.  Alternatively, one could easily construct a candidate set on $\chi$ based on a regular mesh and employ a row exchange algorithm.

\begin{figure}
\begin{center}
\includegraphics[width = 5in]{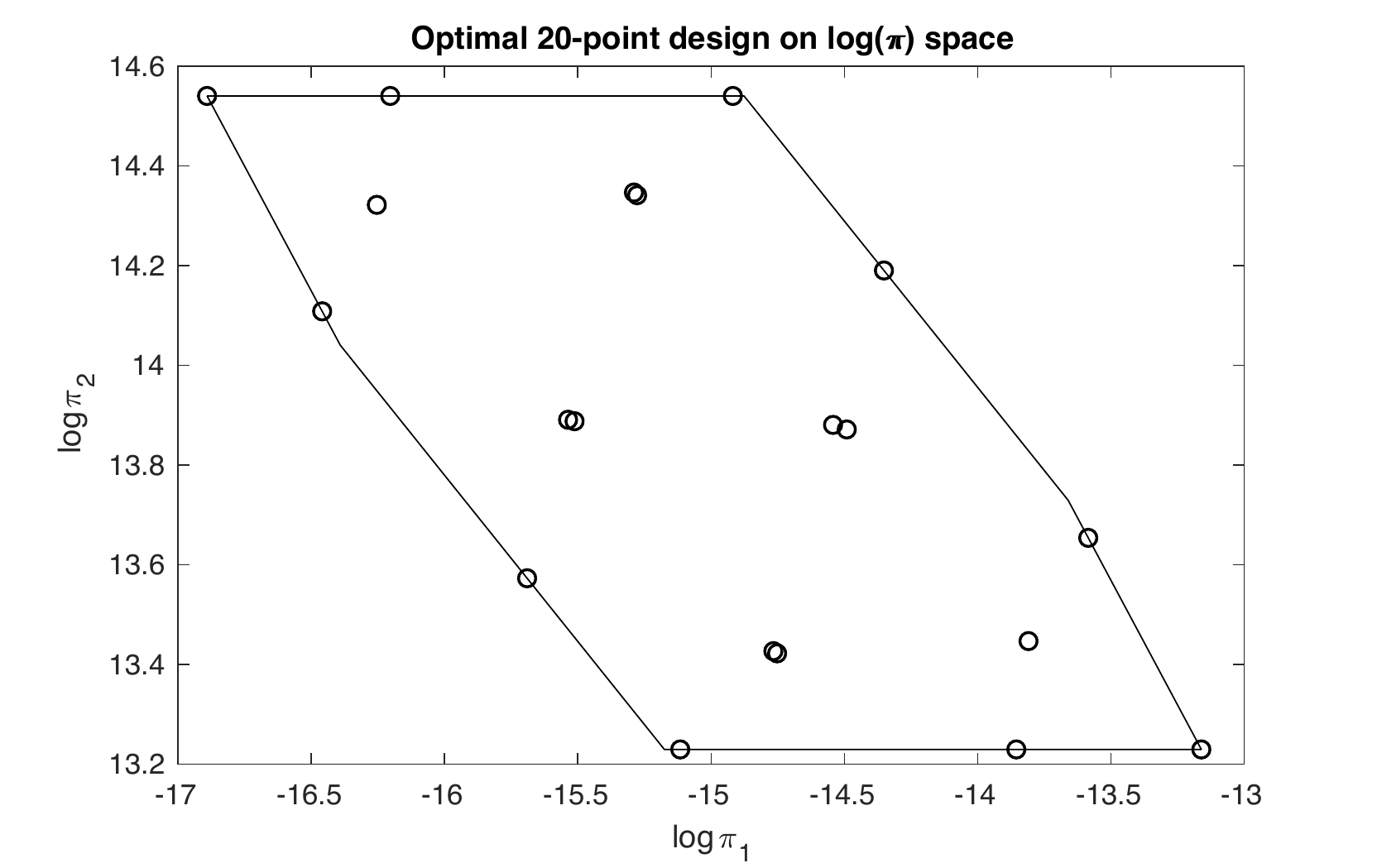}
\caption{Nonparametric design construction using Fast Flexible Filling.}
\label{MV-Optimal2DDesign}
\end{center}
\end{figure}

There are disadvantages associated with either of these approaches.  When a set of uniformly-spaced grid points in $\chi$ are transformed to points in $\chi_{\pi}$ or log$\chi_{\pi}$, the resulting set of points can be highly non-uniform, causing the search routine to miss potentially large subregions of $\chi_{\pi}$ or log$\chi_{\pi}$.  As a simple illustration, a 20$^3 = 8000$-point mesh on $\chi$ for the pump example leads to the irregular grid on log$\chi_{\pi}$ shown in Figure \ref{2DAnd3DDesignSpaces}(d).  In higher-dimensional examples such as the heat-exchanger example discussed in the next section, the problem is exacerbated.  We found that this difficulty can be avoided by using a continuous unidimensional search within the coordinate exchange algorithm.  In the pump and heat-exchange examples discussed herein, we use the L-BFGS-B optimizer (Byrd et al., 1995) from the ``optim'' function in R to perform the continuous unidimensional search.  We used criterion (\ref{MVCriterion}) to construct an optimal design for the pump example with $n = 20$, where the two response models are both full fourth-order models on the log$\chi_{\pi}$ space.  The resulting design is shown in Figure \ref{MV-Optimal2DDesign}.

\subsubsection{Search the $\chi_{\pi}$ space.}  At first glance, it would seem that searching the $\chi_{\pi}$ space directly for the optimal design would be advantageous because the reduction in the size and dimension of the space afforded by the DA model.  We have found that this approach is generally problematic for a number of reasons.  First, the space can be highly irregular, and even non-convex.  Albrecht et al. (2013, Figure 6a) provide one such example; Figure \ref{2DAnd3DDesignSpaces}(b) is another.  In such cases it can be very difficult to accurately articulate the constraints, which are often nonlinear.  If the constraints can be identified, then a continuous optimizer could be used in conjunction with the continuous $\chi_{\pi}$ space to identify the design.  In order to identify the corresponding treatment combinations in $\chi$ (required to actually run the design), basic variable combinations in $\chi$ that lead to the solution design points in $\chi_{\pi}$ must still be identified.  This requires a nonlinear optimization for every design point.  For example, suppose $\boldpi_i$ is the $i$th design point (vector) in $\chi_{\pi}$.  Let $\pi(\boldv_i)$ denote the $\boldpi$ vector that results from the combination of basic variables given by $\boldv_i$.  One approach we have implemented with success is to solve use a continuous optimizer in $\chi$ to find:
\begin{equation}
\boldv_i = \argmin_{\boldv \in \chi}||\pi(\boldv) - \pi_i|| \quad \textrm{for }i = 1,\ldots,n
\label{BackSolve}
\end{equation}
\noindent This approach, while feasible, can add substantially to the computing time required for design construction, especially for large designs.  Because of the above issues, we do not recommend directly searching the $\chi_{\pi}$ space to find the optimal design.

\subsubsection{Search the log$\chi_{\pi}$ space.}  It is straightforward to show that if the $\chi$ is a hyper-rectangle, then the log$\chi_{\pi}$ is a (convex) parallelotope.  See, e.g., \citet{gover}.  Thus it follows that the feasible points form a convex hull bounded by hyperplanes defined by the facets of the convex hull.  Efficient algorithms exist for identifying the facets (and therefore the equations of the bounding hyperplanes)---and for testing whether or not an arbitrary point is contained in the hull---have been developed (Barber et al., 1996).  This suggests that the use of nonlinear optimizers that handles linear constraints (such as generalized reduced gradient algorithms) might be effective for design construction in log$\chi_{\pi}$.  A disadvantage of this approach is that the backsolving for the treatment combinations in $\chi$, as described above in connection with (\ref{BackSolve}) must still be carried out.

\subsubsection{Key findings for parametric designs}

Key points for optimal design construction using parametric models are summarized below.
\begin{enumerate}
\item Search the $\chi$ space, transforming to the log$\chi_{\pi}$ space for evaluation of the design criterion.
\item Use the coordinate exchange algorithm (Meyer and Nachtsheim, 1995) in conjunction with a continuous optimizer when carrying out a coordinate-wise search in $\chi$.
\item The values of the coordinates of the design in the log$\chi_{\pi}$ space can become unduly large or small, leading to numerical instability.  It is often necessary for larger problems such as that discussed in Section 6, to transform the log$\chi_{\pi}$ space to the $[-1,1]^{p-k}$ hypercube for numerical stability.
\end{enumerate}

\subsection{Nonparametric design \label{NonparametricDesign}}
Because of the fact that the log$\chi_{\pi}$ space is a parallelotope, an approximately uniform design can be obtained as follows:

\begin{enumerate}
\item Identify the extreme vertices of the log$\chi_{\pi}$ parallelotope following Barber et al., (1996).
\item Identify the boundaries of the smallest hyper rectangle, denoted log$\chi_{\pi}^R$, containing the log$\chi_{\pi}$ space by identifying the minima and maxima of the extreme vertices.
\item Generate a very large number of random (uniform) points in the log$\chi_{\pi}^R$ hyper rectangle and retain those that fall within the log$\chi_{\pi}$ parallelotope.  The latter can be accomplished using the test described in Barber et al., (1996).  By the rejection method, the resulting points provide a random uniform sample on log$\chi_{\pi}$.
\item Use the Fast Flexible Filling (FFF) method of Lekivetz and Jones (2014) to select an $n$-point subset which comprises the design.  The FFF method is implemented in the JMP statistical software, but only works there with continuous spaces.  We employed a modified version of the algorithm provided by Jones (2018, personal communication) that works with candidate sets.
\end{enumerate}

\begin{figure}
\begin{center}
\includegraphics[width = 6in]{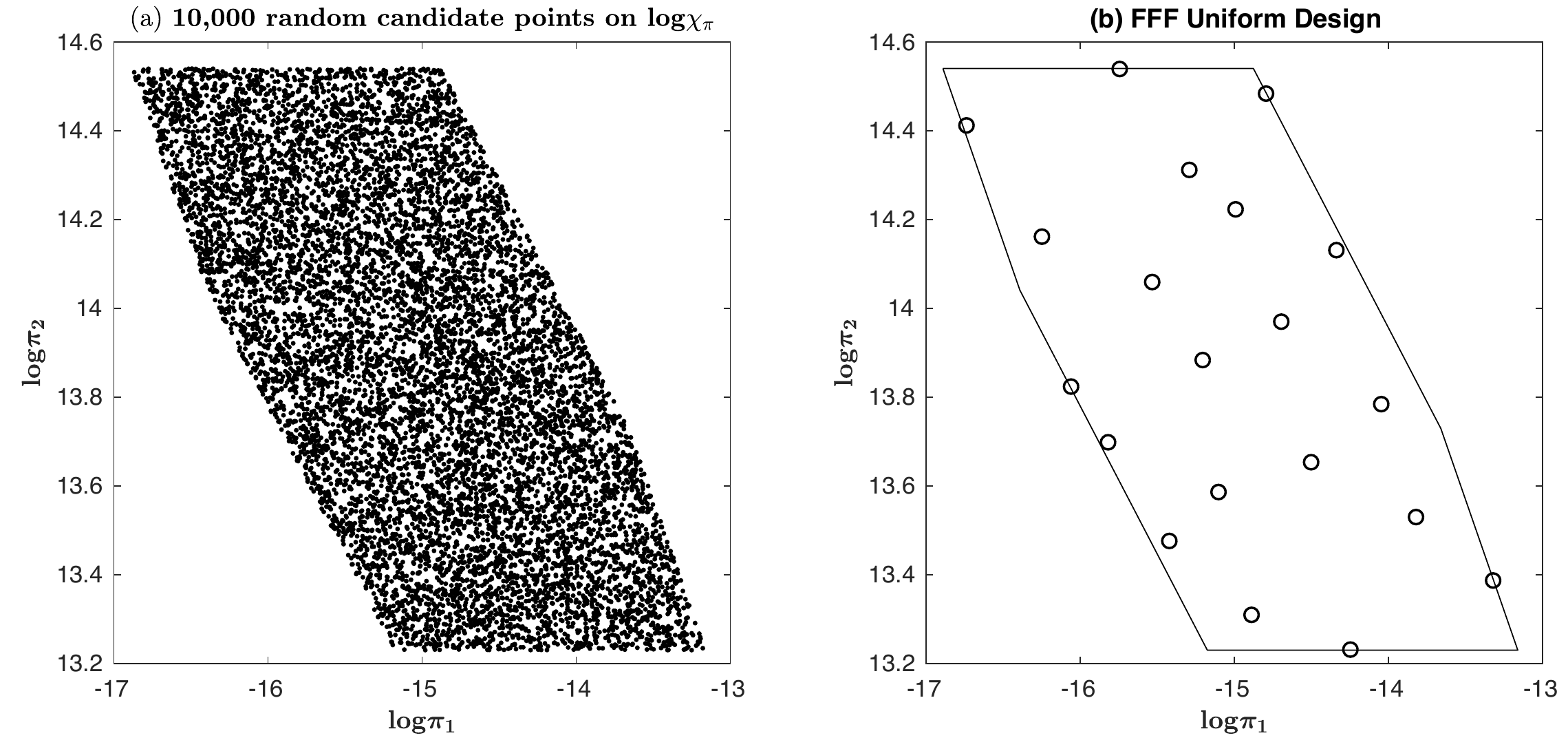}
\caption{Nonparametric design construction using Fast Flexible Filling.}
\label{FFFDesign2D}
\end{center}
\end{figure}

\noindent For the pump example, we generated 10,000 uniformly distributed points on log$\chi_{\pi}$ as shown in Figure \ref{FFFDesign2D}(a).  Applying the FFF method with $n = 20$ to the candidate set in Figure \ref{FFFDesign2D}(a) resulted in the nonparametric design of Figure \ref{FFFDesign2D}(b).

We now turn to design construction for a more complicated example involving nine base variables and five $\pi$ groups. We also construct the ``Robust-DA'' design (Albrecht et al., 2013) for this example.

\section{Heat Transfer Example \label{HeatTransferExample}}

In this section we consider a more complex example involving the design of a heat exchanger.  The model has two responses, and the models for the two responses are not a function of the same sets of dimensionless factors.  We derive the DA model and construct parametric and non-parametric designs using the methodologies described in the preceding section.

\begin{figure}
\begin{center}
\includegraphics[width = 6in]{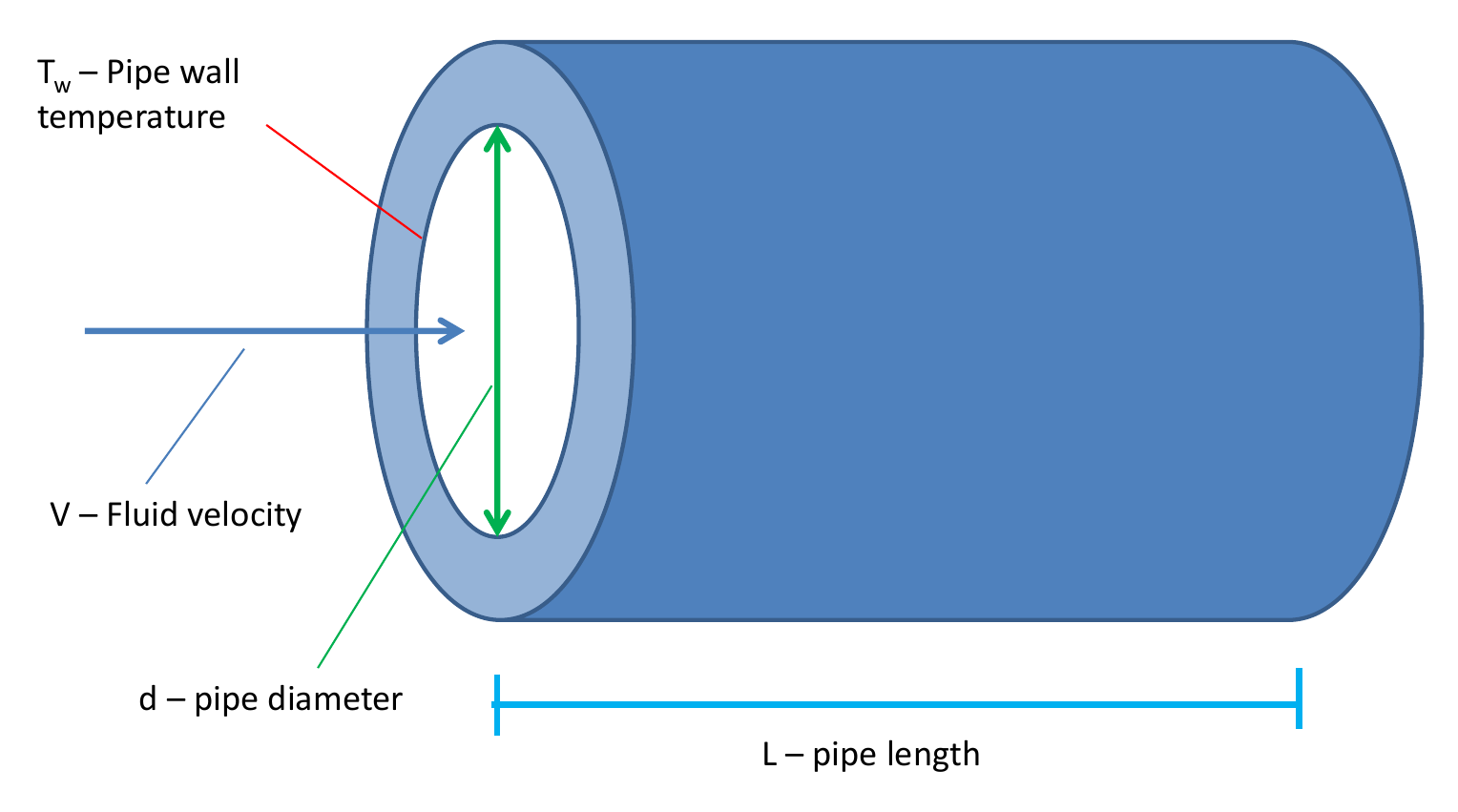}
\caption{Heat exchange schematic.}
\label{HeatExchangerDiagram}
\end{center}
\end{figure}

Assume that engineers wish to design a heat exchanger using circular tubing to extract heat energy at a required rate from a system. In particular, engineers would like to identify the heat exchanger design that minimizes energy lost due to friction between the fluid and the tubing while maximizing the rate of heat extraction from the tubing to the fluid. To achieve the optimal heat exchanger design, engineers can control a number of design parameters such as the length and diameter of the tubing, the average fluid velocity, and the viscosity of the fluid.  These parameters are expected to change both the energy lost due to friction and the rate of heat extraction from the tubing. This system is described in the Figure \ref{HeatExchangerDiagram}.

To create a dimensionless model of the system, we first need to identify all of the base variables that influence friction losses and heat extraction rate. If we assume that the elevation change between the start and end of the pipe is negligible and that the flow through the tubing is fully developed laminar flow to simplify the problem for illustrative purposes, then the pressure loss in the tubing is governed by the following formulas per White (1999):
\begin{align*}
\Delta P &=\rho g h_f \\
h_f&=\frac{f L V^2}{2dg} \\
f&=\frac{64}{\textit{Re}_d} \\
\textit{Re}_d&=\frac{\rho Vd}{\mu}
\end{align*}

\noindent If we also assume a constant inner surface temperature of the tubing to further simplify the problem, the heat transfer rate ($Q$) from the tubing to the fluid can be modeled using the following formulas per Incropera (2007):
\begin{align*}
Q&=hA(T_w-T_f) \\
A&=\textrm{internal pipe surface area} =  \frac{\pi Ld^2}{4} \\
\frac{(hd)}{k}&=Nu_d = 3.66
\end{align*}

Combining the variables from Figure (\ref{HeatExchangerDiagram}) and the governing equations above, we obtain the set of base variables shown in Table \ref{BasicVariablesHeatExchanger}. Here, we have ignored intermediate variables, such as $f$, that are already determined in terms of other variables in the governing equations above.  We use Ipsen's (1960) stepwise method as described in Section \ref{OverviewOfDA} to obtain dimensionless representations of the base variables as shown in Table \ref{Ex2DAModel} of Appendix 2.  The DA model is:
\begin{equation}
\left(
\begin{array}{c}
\frac{\Delta P}{\rho V^2} \\
\frac{Q}{\rho d^2 V^3}
\end{array} \right)= \phi\left(\frac{L}{d},\frac{T_W}{T_f},\frac{\mu}{\rho d V},\frac{gd}{v^2},\frac{KT_f}{\rho D v^3}\right)
\label{HeatExDAModel}
\end{equation}
\noindent The dimension of the design problem has been reduced from nine factors to five factors due to the fact that four basic dimensions are present in the base variables.  We will use $\pi_0^{(1)}$ and $\pi_0^{(2)}$ to denote the two dimensionless response variables in (\ref{HeatExDAModel}), and $\pi_1,\ldots,\pi_5$ to denote the five design factors, respectively.  It is important to note that the first response, $\pi_0^{(1)}$, is dependent on all five of the dimensionless factors, whereas $\pi_0^{(2)}$ depends only on $\pi_1$, $\pi_3$, and $\pi_4$.  Thus we write:
\begin{align}
\begin{split}
\pi_0^{(1)} &= \phi_1(\pi_1,\pi_2,\pi_3,\pi_4,\pi_5) \\
\pi_0^{(2)} &= \phi_2(\pi_1,\pi_3,\pi_4)
\end{split}
\end{align}


We noted in Section 2 that eliminating dimensions in different sequences can lead to alternative sets of dimensionless variables, and that engineers prefer $\pi$ groups that correspond to dimensionless quantities whose properties are well known.  In (\ref{HeatExDAModel}), $\pi_0^{(1)}$ and $\pi_3$ correspond to the Euler number, and Reynolds number, respectively.

\begin{table}
\caption{Basic variables for the heat exchanger example}
\begin{center}
\begin{tabular}{l|c|c|c}
Base variable & Symbol & SI Units & Base Quantity \\
\hline
Pressure change (response) & $\Delta P$ & $Pa$ & $M/Lt^2$ \\
Heat transfer rate (response) & $Q$ & $W$ & $ML^2/t^3$ \\
\hline
Inner diameter of pipe & $d$ & $m$ & $L$ \\
Length of pipe & $L$ & $m$ & $L$ \\
Mean velocity of fluid & $V$ & $m/s$ & $L/t$\\
Temperature of inner pipe wall & $T_w$ & $K$ & $T$ \\
Mean temperature of fluid in the pipe & $T_f$ & $K$ & $T$ \\
Viscosity of fluid & $\mu$ & $N-s/m^2$ & $M/Lt$ \\
Density of fluid & $\rho$ & $kg/m^3$ & $M/L^3$ \\
Gravitational acceleration & $g$ & $m/s^2$ & $L/t^2$ \\
Thermal conductivity of fluid & $K$ & $W/(m-K)$ & $ML/Tt^3$ \\
\hline
\end{tabular}
\end{center}
\label{BasicVariablesHeatExchanger}
\end{table}%

\subsection{Parametric Design}
\begin{figure}
\begin{center}
\includegraphics[width = 5in]{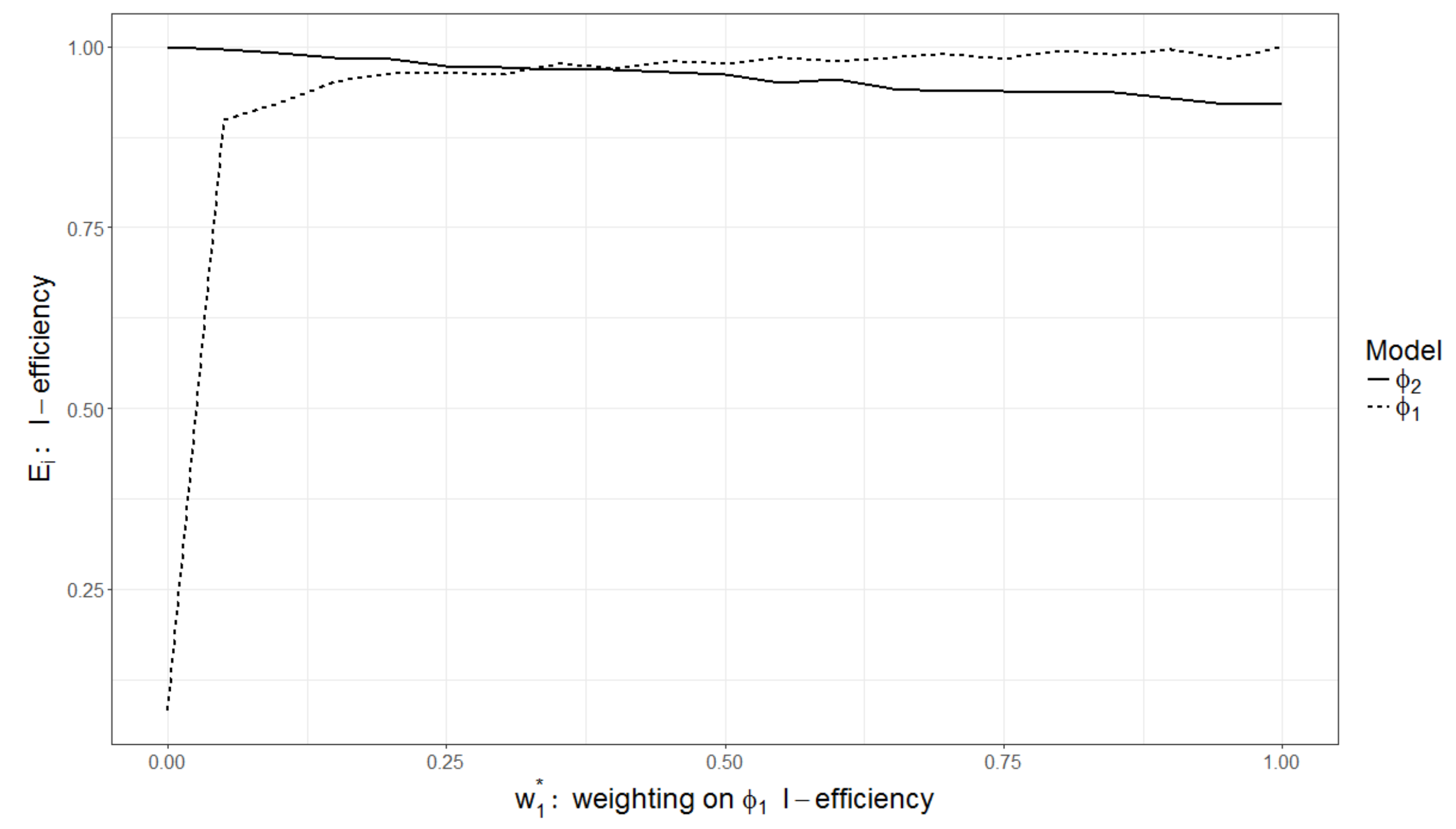}
\caption{I-efficiencies of the optimal design for as a function of $w^*_1$}
\label{ModelRobustPlot}
\end{center}
\end{figure}

\begin{figure}
\begin{center}
\includegraphics[width = 5in]{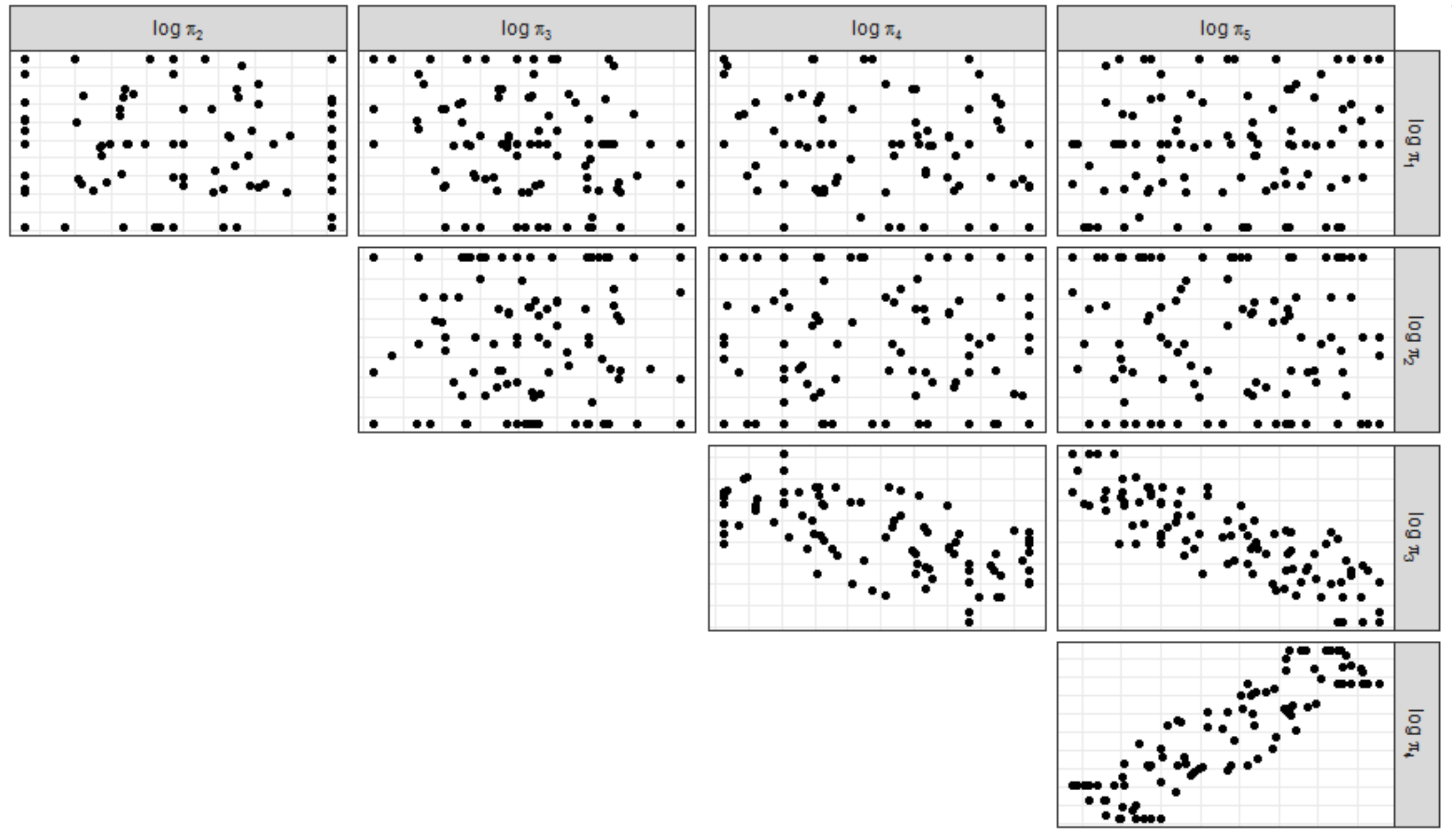}
\caption{Two-dimensional projections of the the optimal design for $w^*_1 = 1$}
\label{Projections5piDesign}
\end{center}
\end{figure}

We first use multivariate design criterion (\ref{MVCriterion}) to construct designs for a full third-order model with $n = 100$.  Apart from a constant, the criterion is a linear combination of the I-optimality criterion values for the two models where the weights are given by the variances $w_i^{-1} = \sigma_i^2$ of the responses.  Often this information will not be available.  In what follows, we reparameterize the weights as $w_i^* = w_i^{-1}/(w_1^{-1}+w_2^{-1})$ for $i = 1,2$, so that the weights sum to one. With $\boldw^* = (w_1^*,w_2^*)'$ rewrite (\ref{MVCriterion}) as:
\begin{equation}
I_{MV}(\xi_n,\boldw^*) \propto \sum_{i=1}^2w_i^*\textrm{Trace}[\boldD_i\boldM_i]
\end{equation}
We then compute optimal designs $\xi_n(w_1^*)$ for $0 \le w_1^* \le 1$ in an effort to identify a design that is efficient for both models, irrespective of the weights. Let $\boldD_i(\xi_n)$ denote the dispersion matrix of nonsingular design $\xi_n$ for model $i$, and let $\xi_n^*$ denote an I-optimal design for model $i$, for $i = 1,2$.  The I-efficiency of a design $\xi_n$ for model $i$ is:
\[
\textrm{E}_i(\xi_n) = \frac{\textrm{Trace}[\boldD_i(\xi_n^*)\boldM_i]}{\textrm{Trace}[\boldD_i(\xi_n)\boldM_i]}
\]
Plots of $\textrm{E}_i[\xi_n(w_1^*)]$ for $i = 1,2$ as a function of $w_1^*$ are provided in Figure \ref{ModelRobustPlot}.  The maximin design results with $w_1^* =0.35$, in which case the I-efficiency for both models is approximately 97\%.  Alternatively, if we choose to work with the optimal design for $\phi_1$ (i.e., the larger model), the efficiency for estimation of $\phi_2$ is still 92\%.  This robustness finding is similar to the model robustness result in Cook and Nachtsheim (1982) which indicated that designing for the highest degree polynomial when the model space consists of a set of nested polynomials is an effective strategy.  A plot of the two-dimensional projections for the optimal design for $w^*_1 = 1$ is shown in Figure \ref{Projections5piDesign}.

\subsection{Nonparametric (uniform) design}

\begin{figure}
\begin{center}
\includegraphics[width = 5in]{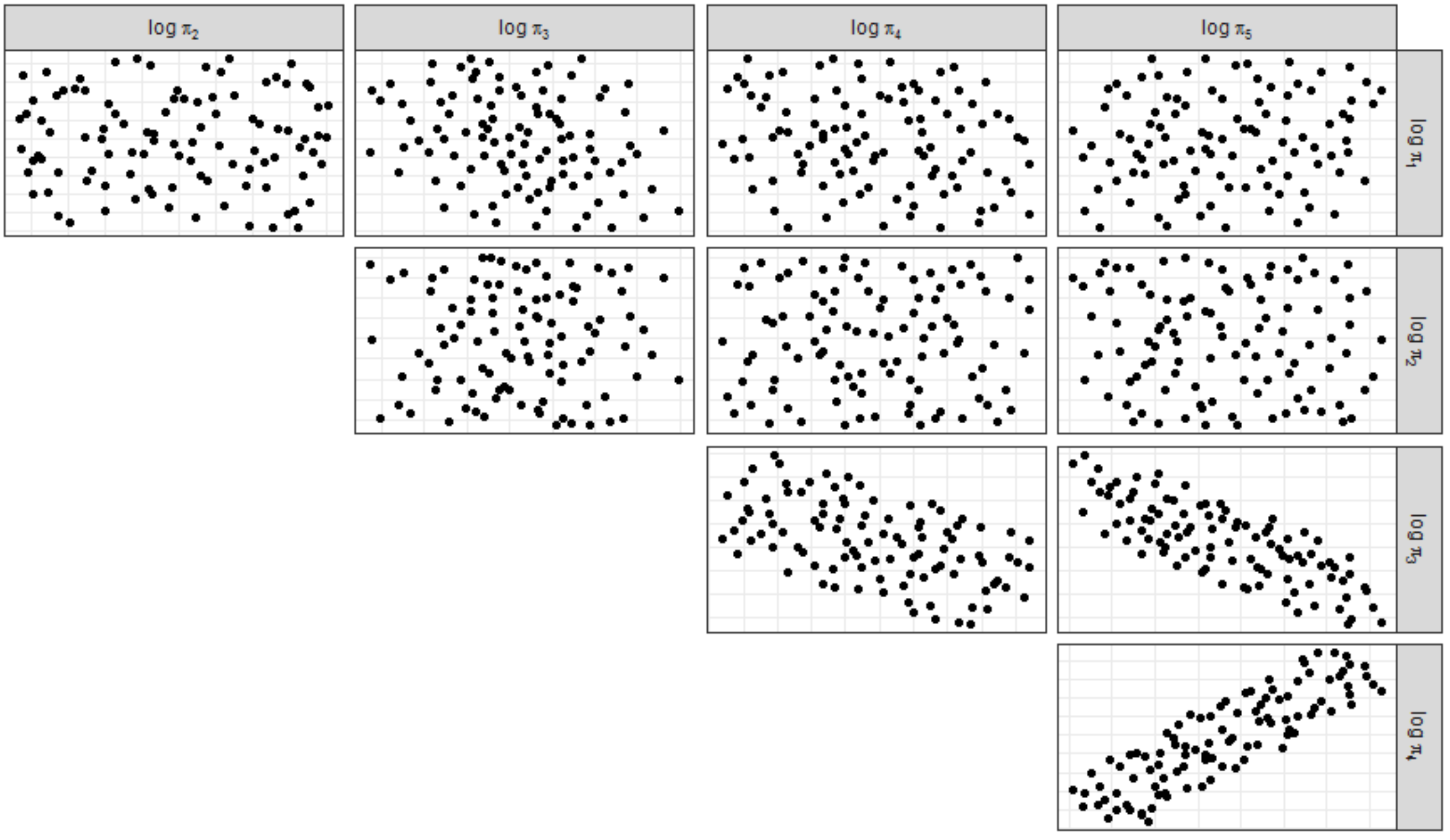}
\caption{Two-dimensional projections of the the uniform design}
\label{UniformProjections}
\end{center}
\end{figure}

We next employ the four-step procedure of Section \ref{NonparametricDesign} to construct an approximately uniform design on log$\chi_{\pi}$.  As before we generate 10,000 uniformly distributed points in log$\chi_{\pi}$ using the rejection method.  We note that for this problem, log$\chi_{\pi}$ occupied about 8\% of the minimum enclosing hyperrectangle so that about 92\% of points generated uniformly on log$\chi_{\pi}$ were rejected.  We then used the FFF method to obtain a subset of $n=100$ design points.   Two-dimensional projection plots for this design are shown in Figure \ref{UniformProjections}.

How one might choose between the parametric and nonparametric designs of Figures \ref{Projections5piDesign} and \ref{UniformProjections} is not obvious.  Comparing the figures visually, it is clear that the parametric approach places many of its points on the boundaries of the design space.  The uniform design clearly spreads point more evenly throughout the design space.  If the experimenter is undecided as to choice of the parametric or nonparametric approaches, it can be helpful to compute the $I$-efficiency of the uniform design.  In this case the efficiency of the uniform design, relative to the design for $\phi_1$ shown in Figure \ref{Projections5piDesign} is 29\%.

As noted above, in order to run the design we still need to identify the treatment combinations in $\chi$ that lead to the design points in log$\chi_{\pi}$, and we use the optimization procedure identified in (\ref{BackSolve}) to do so. Of course, for any design vector $\boldpi_i$, there is a collection of treatments $\boldv_i$ that lead to $\pi(\boldv_i) = \mathbf{\pi}_i$.  This motivated Albrecht et al., (2013) to propose DA-robust designs, which give good designs for the DA model which are also highly efficient for an empirical design in $\chi$.  We consider this approach next.

\subsection{Robust-DA design \label{RobustDA}}

\begin{figure}
\begin{center}
\includegraphics[width = 5in]{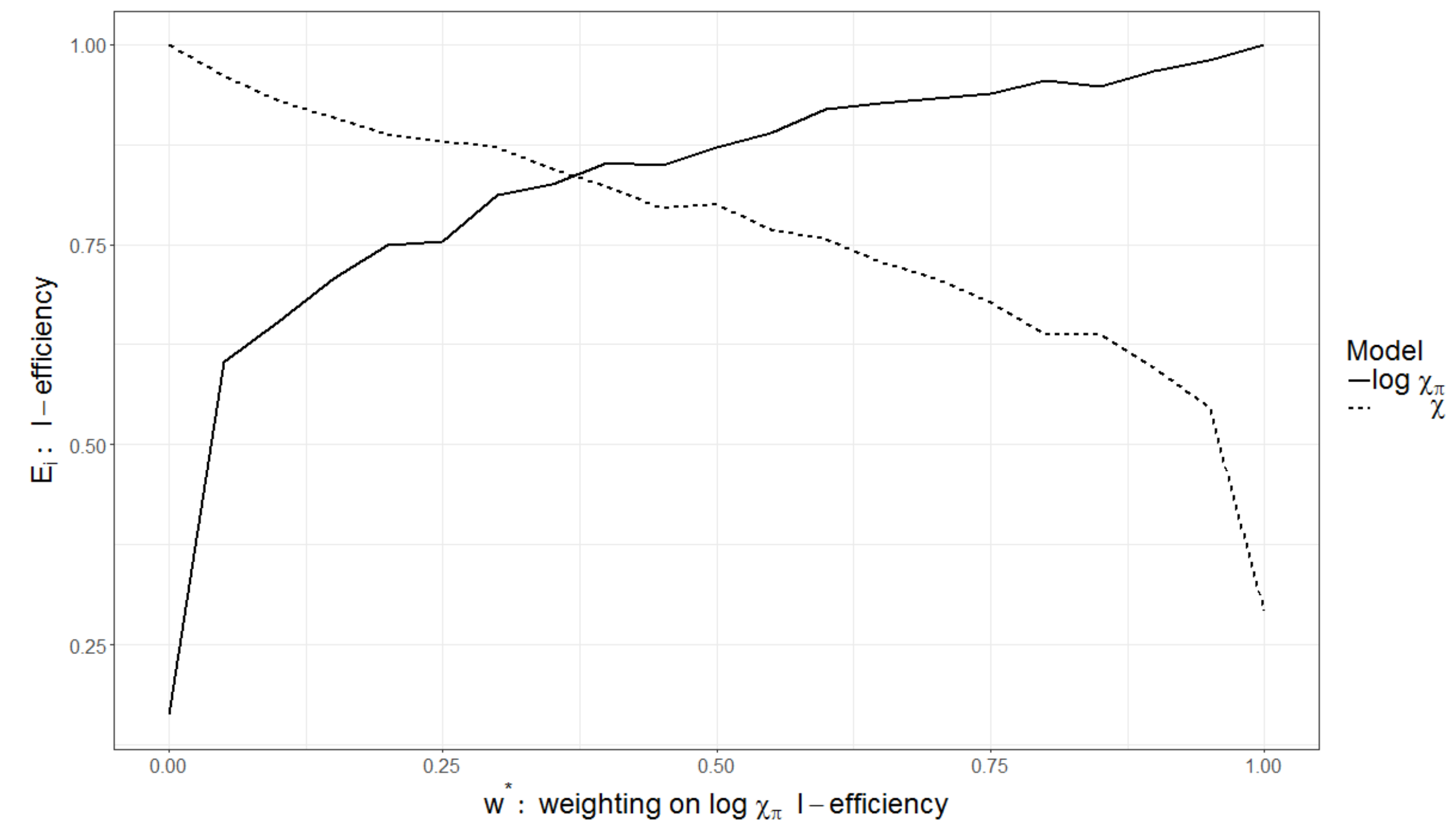}
\caption{Identifying the maximin DA Robust design}
\label{RobustDAPlot}
\end{center}
\end{figure}

\begin{figure}
\begin{center}
\includegraphics[width = 5in]{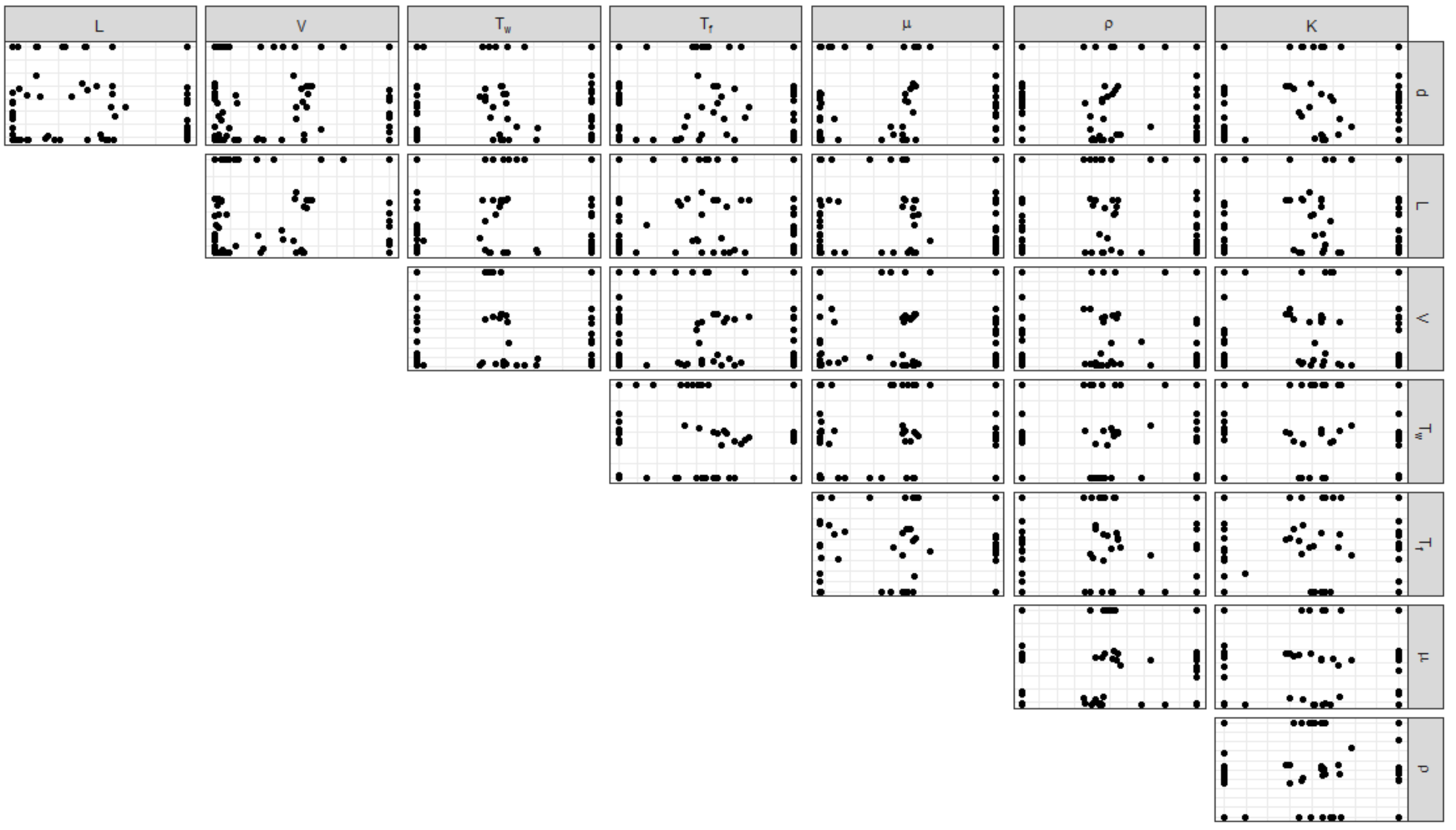}
\caption{Two-dimensional projections of the the Robust DA design on $\chi$}
\label{RobustDAChiProjections}
\end{center}
\end{figure}

\begin{figure}
\begin{center}
\includegraphics[width = 5in]{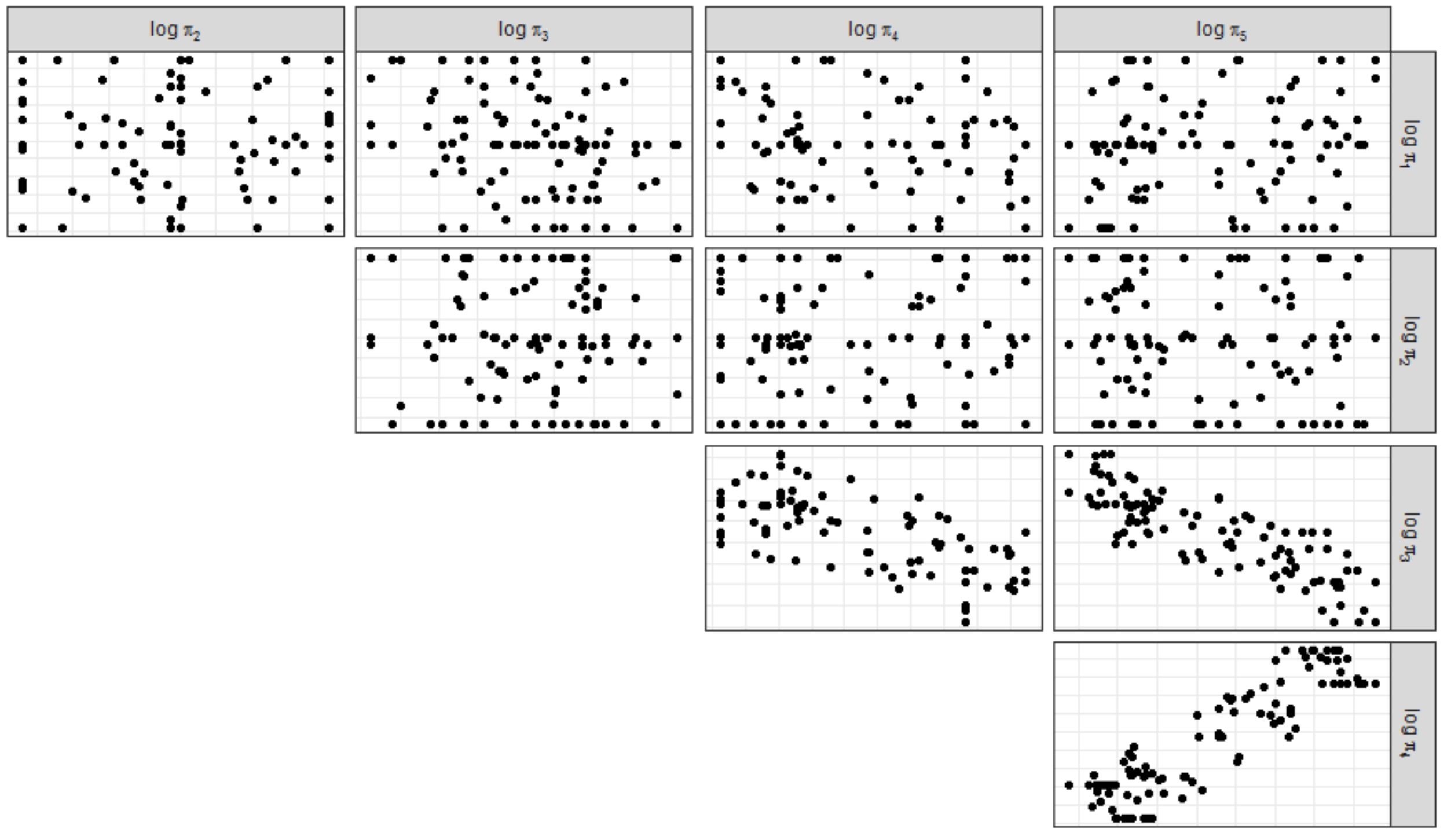}
\caption{Two-dimensional projections of the the Robust DA design on log$\chi_{\pi}$}
\label{RobustDAPiProjections}
\end{center}
\end{figure}

As noted by Albrecht et al., (2013), the omission of an important, active dimensionless factor can be fatal to the DA process.  In an effort to salvage the results of such an experiment, they suggested the Robust-DA approach, which produces efficient ``empirical'' designs in the $\chi$ space, while simultaneously yielding efficient designs in the log$\chi_{\pi}$ space.  For any design $\xi_n$ on $\chi$, let E$_{\chi}(\xi_n)$ denote the I-efficiency of the design for the full quadratic model in the $\chi$ space, and let E$_{\textrm{log}\chi_{\pi}}(\xi_n)$ denote the efficiency of the design for the model of interest (i.e., full third- or higher-order model) on log$\chi_{\pi}$.  The compound criterion considered by Albrecht et al., (2013) is:
\[
c(\xi_n,w) = wE_{\textrm{log}\chi_{\pi}}(\xi_n) + (1-w)E_{\chi}(\xi_n)
\]
\noindent where $0 \le w \le 1$.  Let $\xi_n^w$ denote the design that optimizes $c(\xi_n,w)$ for fixed $w$.  Here we search for the maximin design:
\[
\xi_n^{w^*} = \max_w \min[E_{\textrm{log}\chi_{\pi}}(\xi_n^{w}),E_{\chi}(\xi_n^{w})]
\]
\noindent In Figure \ref{RobustDAPlot}, we plot the values of $E_{\textrm{log}\chi_{\pi}}(\xi_n^{w})$ and $E_{\chi}(\xi_n^{w})$ for a grid of $w$ values between zero and one.  The maximin design occurs for $w^* =0.35$, in which case the efficiencies for the DA model and the empirical model are 83\% and
85\%, respectively.  The two-dimensional projections for the empirical design and the two-dimensional projections of the DA design in the log$\chi_{\pi}$ space are shown in Figures \ref{RobustDAChiProjections} and \ref{RobustDAPiProjections}, respectively.  We note that the two-dimensional projections of the maximin design in $\chi$ clearly reflect the patterns in I-optimal response surface designs, having points either on the border or near the center of the design space.

\section{Discussion}
In this paper, we have developed new methodology for designing DA experiments
when there is more than a single response.  We began by extending the
Buckingham $\Pi$-Theorem to the multivariate case.  We then developed basic
criteria for multivariate design of experiments and we illustrated various
approaches for a DA problem involving mechanical pump design and another
involving the design of a heat exchanger.  This methodology is applicable when
the dimension matrix for the response variables is contained in the span of
dimension matrix for the explanatory variables \eqref{functional-form1}.
Absence of a key fundamental dimension is problematic but such a design flaw
can be mitigated through close collaboration with domain experts and or the
Robust-DA design approach provided by Albrecht et al. (2013), as demonstrated
in Section \ref{RobustDA}.  Through the incorporation of DA into the
multivariate design of experiments framework, we can design cheaper
experiments by reducing the dimension of the design space and by developing a
setting for which scalable experiments can be performed.

Some interesting issues arise in the construction of the optimal designs, and we explored three alternative approaches in Section \ref{DesignIssues}.  Our numerical experience suggests that the most efficient search procedure results from use of the coordinate exchange algorithm on $\chi$, the design space of the orginal variables, with an embedded, one-dimensional continuous optimizer.  R code for computing the optimal parametric, and uniform non-parametric designs is available on the journal website.

\pagebreak

\section{Appendix 1}

In this Appendix, we provide the proofs of the Multivariate
Buckingham $\Pi$-Theorem and then its Corollary. The following is the proof of
the Multivariate Buckingham $\Pi$-Theorem.

\begin{proof}
We have $\boldY \in \mathbb{R}^r$ as our response vector,
$\boldx \in \mathbb{R}^p$ as our vector of predictors, and $k$ fundamental
dimensions where assumption (ii) states that all fundamental dimensions are
represented by elements in the vector of predictors. The dimensions of
elements in either the response vector or the vector of predictors can be
written as
\begin{align*}
 \left[Y_j\right] &= \prod\limits_{i=1}^k L_i^{a_{ij}}, \quad j = 1,..., r, \\
 \left[x_j\right] &= \prod\limits_{i=1}^k L_i^{b_{ij}}, \quad j = 1,..., p.
\end{align*}
Now, for the first dimension $L_1$, consider invariance under arbitrary
scaling. Let $L_1^* = e^{\varepsilon}L_1$ where $\varepsilon \in \mathbb{R}$
and according to this scaling define
\begin{equation}
\label{Y-scaling}
  Y_i^* = e^{a_{1i}\varepsilon}Y_i, \quad i = 1,..., r,
\end{equation}
\begin{equation}
\label{X-scaling}
  x_j^* = e^{b_{1j}\varepsilon}x_j, \quad j = 1,..., p.
\end{equation}
These equations define a one-parameter Lie group of the $p + r$ quantities $(x_1,...,x_p,Y_1,...,Y_r)$. This group is induced by the one-parameter group of scalings of the fundamental dimension $L_1$. Assumption (iv) says that equation $(\ref{functional-form1})$ holds iff
$$
  (Y_1^*,...,Y_r^*)^{\prime} = \boldf(x_1^*,...,x_p^*)
$$
holds for all $\varepsilon \in \mathbb{R}$. Consider the following three cases that occur when trivialities exist in our original problem.
\begin{enumerate}
\item[(i)]: $b_{11} = \cdots = b_{1p} = 0$ and/or at least one
  $a_{1j} \neq 0$ for some $j = 1,...,r$ which implies that $L_1$ is not a
  fundamental dimension for the problem and $\boldY_j = 0$ whenever
  $a_{1j} \neq 0$.
\item[(ii)]: If in case (i) we have $a_{1j} \neq 0$ for all $j = 1,...,r$
  then $\boldY = \boldZero_r$ where $\boldZero_r$ is the 0's vector in
  $\mathbb{R}^r$.
\item[(iii)]: If only one $b_{1j} \neq 0$ for some $j = 1,..., p$ and
  $a_{1i} = 0$ for all $i = 1,..., r$, then either $\boldY = \boldZero_r$
  and $L_1$ is a fundamental dimension for the problem or $\boldY$ is
  independent of $x_j$ and $L_1$ is not a fundamental dimension for the
  problem.
\end{enumerate}
Suppose the problem is set up so that cases (i)-(iii) do not occur. It follows that $b_{1j} \neq 0$ for some $j = 1,...,p$.  Without loss of generality, assume that $b_{11} \neq 0$. Define new measurable quantities
\begin{align*}
  W_{i-1} &= x_i x_1^{-b_{1i}/b_{11}} \quad i = 2,..., p, \\
  W_p &= x_1,
\end{align*}
and
$$
  \boldV = (Y_1 W_p^{-a_{11}/b_{11}},...,Y_r W_p^{-a_{1r}/b_{11}})^{\prime}.
$$
Then formula $(\ref{functional-form1})$ is equivalent to
$
  \boldV = \boldF(W_1,...,W_p)
$
for an unknown function $\boldF$. The group of transformations seen in (\ref{Y-scaling}) and (\ref{X-scaling}) yield
\begin{align*}
  \boldV^* &= \boldV, \\
  W_i^* &= W_i, \quad i = 1,..., p-1, \\
  W_p^* &= e^{b_{11}\varepsilon}W_p,
\end{align*}
so that $(V_1,...,V_r,W_1,...,W_{p-1})$ are invariants of (\ref{Y-scaling}) and (\ref{X-scaling}). Therefore
$
  \boldV^* = \boldF(W_1,...,W_p)
$
holds and
$
  \boldV^* = \boldF(W_1,...,W_{p-1}, e^{\varepsilon b_{11}}W_p)
$
holds for all $\varepsilon \in \R$ as a result of assumption (iv). Consequently $\boldF$ is independent of $W_p$. Moreover, the measurable quantities $(W_1,...,W_{p-1})$ and the elements of $\boldV$ are power-law combinations of the original $(x_1,...,x_p)$. Formula (\ref{functional-form1}) reduces to
$$
  \boldV = \boldH(W_1,...,W_{p-1}),
$$
where all variables are dimensionless with respect to $L_1$ and $\boldH$ is an unknown function. This argument is repeated for the other $k-1$ fundamental dimensions. The repetition of this argument reduces (\ref{functional-form1}) to a dimensionless formula one fundamental dimension at a time. We arrive at the functional form
$$
  \tilde{\boldpi} = \boldh(\pi_1,..., \pi_{p-k}),
$$
where
$$
  \tilde{\boldpi} = \left(\begin{array}{c}
      \tilde{\pi}_1 \\
      \vdots \\
      \tilde{\pi}_r
  \end{array}\right)  =
  \text{diag}\left(\prod_{i=1}^p\boldx_i^{y_{i1}},...,\prod_{i=1}^p\boldx_i^{y_{ir}}\right)
    \left(\begin{array}{c}
      Y_1 \\
      \vdots \\
      Y_r
  \end{array}\right).
$$
Next it is shown that the number of measurable dimensionless predictors is in fact $p - \text{rank}(\boldB)$. This follows immediately since
$$
  \left[ \prod_{j=1}^p x_i^{\pi_{ij}} \right] = 1 \quad \text{if and only if} \quad \boldB x_i = 0
$$
and $\boldB\boldx = 0$ has $p - \text{rank}(\boldB)$ linearly independent solutions. The vectors $\boldy_i$ are chosen such that
$$
  \left[ Y_i \prod_{j=1}^r x_j^{\rho_{ji}} \right] = 1.
$$
This choice is valid because of assumption (ii). Therefore $\boldB\boldy_i = -\bolda_i$ for $i = 1,...,r$ and this completes the proof.
\end{proof}

The following is the proof of the Corollary to the Multivariate Buckingham $\Pi$-Theorem.

\begin{proof}
Conclusion (i) follows using the same techniques in the proof of the Multivariate Buckingham $\Pi$-Theorem. Now to show that conclusion (ii) holds. The argument used to show that conclusion (i) holds shows that variables corresponding to the dimension matrix $\boldC$ can be made into $k^{\prime} = p + r_2 - \text{rank}(\boldC)$ dimensionless quantities. From the proof of the Multivariate Buckingham $\Pi$-Theorem, we have that $p - \text{rank}(\boldB)$ of these dimensionless quantities are predictors. Therefore $k^{\prime} - (p - \text{rank}(\boldB)) = r_2 - \text{rank}(\boldC) + \text{rank}(\boldB)$ are responses. There are $r_1 - r_2$ remaining responses that can be made dimensionless using the dimension matrix $\boldC$. This leaves us with $r_3$ dimensionless responses in total. We can see that a function $\boldg$ exists (satisfying (\ref{functional-form2})) by combining what has already been proved, assumption (i), and the Multivariate Buckingham $\Pi$-Theorem. The other assumptions are necessary for these calculations to hold. This completes the proof.
\end{proof}

To see when the Corollary 1 is applicable and Theorem 1 is not, consider the
following example. Suppose there are three responses and
three predictors where each has fundamental dimensions given by
$\left[Y_1\right] = ML$, $\left[Y_2\right] = ML$, $\left[Y_3\right] = M$,
$\left[x_1\right] = L$, $\left[x_2\right] = T$, $\left[x_3\right] = LT$.
In this setup
$$
  \boldA = \left(\begin{array}{ccc}
  1 &  1 & 1 \\
  1 &  1 & 0 \\
  0 &  0 & 0
  \end{array}\right), \qquad \boldB = \left(\begin{array}{ccc}
  0 &  0 &  0 \\
  1 &  0 &  1 \\
  0 &  1 &  1
  \end{array}\right).
$$
We see that $\boldA \not\subseteq$ span$(\boldB)$. However, we have that
$\left[Y_1\right] = \boldA_1 \subseteq$ span$(\boldA_{-1},\boldB)$.
We can create the new dimensionless response $Y_1^{\prime} = Y_1/Y_2$.
We also have that
$\left[Y_3\right] = \boldA_3 \subseteq$ span$(\boldA_{-3},\boldB)$.
We can create the new dimensionless response $Y_2^{\prime} = Y_1/(Y_3x_1)$.
From here we can use Theorem 1 and consider the dimensionless predictor
$\pi_1 = x_3/(x_1x_2)$ so that the model under consideration is now
$$
  \left(\begin{array}{c}
    Y_1^{\prime} \\
    Y_2^{\prime}
  \end{array}\right) = \boldg(\pi_1).
$$
Our DA model consists of $p - \text{rank}(B) = 1$ dimensionless predictor,
as given by Theorem 1, and two responses.

\pagebreak
\section{Appendix 2: Derivation of heat Exchanger DA model}

The derivation of the heat exchanger DA model, using Ipsen's (1960) stepwise approach, is detailed in Table \ref{Ex2DAModel}.

\begin{table}[h!]
\begin{center}
\begin{tabular}{|ll|cc|cc|cc|cc|}
\hline
&&\mc{2}{|c|}{Step 1 Result:} & \mc{2}{|c|}{Step 2 Result:} & \mc{2}{|c|}{Step 3 Result:} & \mc{2}{|c|}{Step 4 Result:} \\
&& \mc{2}{|c|}{Remove $M$ from} & \mc{2}{|c|}{Remove $L$ from} & \mc{2}{|c|}{Remove $t$ from} & \mc{2}{|c|}{Remove $T$ from}\\
\mc{2}{|c|}{Step 0: Initialize}& \mc{2}{|c|}{Step 0 using $\rho$} & \mc{2}{|c|}{Step 1 using $d$} & \mc{2}{|c|}{Step 2 using $V/d$} & \mc{2}{|c|}{Step 3 using $T_f$} \\
\hline
Variable & Dimension & Var. & Dim. & Var. & Dim. & Var. & Dim. & Var. & Dim. \\
\hline
$\Delta P$	& $ML^{-1}t^{-2}$ & $\Delta P/\rho$ & $t^{-2}L^2$ & $\Delta P/(\rho d^2)$ & $t^{-2}$ & $\Delta P/(\rho V^2)$ & $1$ & $\Delta P/(\rho V^2)$ & $1$ \\

$Q$	& $mL^2t^{-3}$ & $Q/\rho$	& $t^{-3}L^5$  & $Q/(\rho d^5)$	& $t^{-3}$  & $Q/(\rho d^2 V^3)$	& $1$ & $Q/(\rho d^2 V^3)$	& $1$\\
\hline
$d$ & $L$ & $d$ & $L$ & \mc{2}{|c|}{\text{------------------}} & \mc{2}{|c|}{\text{------------------}} & \mc{2}{|c|}{\text{------------------}} \\

$L$	& $L$ & $L$	& $L$ & $L/d$	& $1$ & $L/d$	& $1$ & $L/d$	& $1$\\

$V$ & $Lt^{-1}$ & $V$ & $Lt^{-1}$ & $V/d$ & $t^{-1}$  &  \mc{2}{|c|}{\text{------------------}} & \mc{2}{|c|}{\text{------------------}} \\

$T_W$	& $T$ & $T_W$	& $T$ & $T_W$	& $T$ & $T_W$	& $T$ & $T_W/T_f$	& $1$ \\

$T_f$	& $T$ & $T_f$	& $T$ & $T_f$	& $T$ & $T_f$	& $T$ &  \mc{2}{|c|}{\text{------------------}}  \\

$\mu$	& $ML^{-1}t^{-1}$ & $\mu/\rho$	& $L^2t^{-1}$ & $\mu/(\rho d^2)$	& $t^{-1}$ & $\mu/(\rho dV)$ & $1$  & $\mu/(\rho dV)$ & $1$\\

$\rho$ & $ML^{-3}$ & \mc{2}{|c|}{\text{------------------}} & \mc{2}{|c|}{\text{------------------}}  & \mc{2}{|c|}{\text{------------------}} & \mc{2}{|c|}{\text{------------------}}  \\

$g$ & $Lt^{-2}$ & $g$ & $Lt^{-2}$ & $g/d$ & $t^{-2}$  & $gd/v^2$ & $1$ & $gd/v^2$ & $1$  \\

$K$ & $MLt^{-1}t^{-3}$ & $K/\rho$ & $L^4T^{-1}t^{-3}$ & $K/(\rho d^4)$ & $T^{-1}t^{-3}$  & $K/(\rho dV^3)$ & $T^{-1}$ & $KT_f/(\rho dV^3)$ & $1$  \\

\hline
\end{tabular}
\caption{Deriving the DA model for Example 2}
\label{Ex2DAModel}
\end{center}
\end{table}

\end{document}